\documentclass[useAMS,usenatbib,a4paper]{mn2e}
\usepackage{amsmath,amsfonts,latexsym,graphicx,amssymb,times,appendix,subfigure,bm,multirow,color}
\newcommand{\bsf}[1]{\textbf{\textsf{#1}}}
\newcommand{\slfrac}[2]{\left.#1\middle/#2\right.}
\newcommand{\sig}{\textrm{sig}}
\newcommand{\SN}{\textsc{SkyNet}}

\graphicspath{{Figs/}}

\title[\SN]{\SN: an efficient and robust neural network training tool for
  machine learning in astronomy} \author[P.~Graff et al.]{Philip Graff$^1$\thanks{Email:
    \texttt{philip.b.graff@gmail.com}}, Farhan Feroz $^2$\thanks{Email:
    \texttt{f.feroz@mrao.cam.ac.uk}}, Michael P. Hobson$^2$ and
  Anthony Lasenby$^{2,3}$\\ $^1$Gravitational Astrophysics Laboratory,
  NASA Goddard Space Flight Center, 8800 Greenbelt Rd., Greenbelt, MD
  20771, USA\\ $^2$Astrophysics Group, Cavendish Laboratory, JJ
  Thomson Avenue, Cambridge CB3 0HE, UK\\ $^3$Kavli Institute for
  Cosmology, Madingley Road, Cambridge CB3 0HA, UK} 
\date{Accepted ---. Received ---; in original form \today}

\pagerange{\pageref{firstpage}--\pageref{lastpage}} \pubyear{2013}

\begin{document}

\label{firstpage}

\maketitle

\begin{abstract}
We present the first public release of our generic neural network
training algorithm, called {\sc SkyNet}. This efficient and robust
machine learning tool is able to train large and deep feed-forward
neural networks, including autoencoders, for use in a wide range of
supervised and unsupervised learning applications, such as regression,
classification, density estimation, clustering and dimensionality
reduction. {\sc SkyNet} uses a `pre-training' method to
obtain a set of network parameters that has empirically been shown to be close to a good solution, 
followed by further optimisation
using a regularised variant of Newton's method, where the level of regularisation is determined
and adjusted automatically; the
latter uses second-order derivative information to improve
convergence, but without the need to evaluate or store the full
Hessian matrix, by using a fast approximate method to calculate
Hessian-vector products. This combination of methods allows for the
training of complicated networks that are difficult to optimise using
standard backpropagation techniques. {\sc SkyNet} employs convergence
criteria that naturally prevent overfitting, and also includes a fast
algorithm for estimating the accuracy of network outputs. The utility
and flexibility of {\sc SkyNet} are demonstrated by application to a
number of toy problems, and to astronomical problems focusing on the
recovery of structure from blurred and noisy images, the
identification of gamma-ray bursters, and the compression and
denoising of galaxy images. The {\sc SkyNet} software, which is
implemented in standard ANSI C and fully parallelised using MPI, is
available at {\tt http://www.mrao.cam.ac.uk/software/skynet/}.
\end{abstract}


\section{Introduction}\label{sec:Intro}

In modern astronomy, one is increasingly faced with
the problem of analysing large, complicated and multidimensional data
sets. Such analyses typically include tasks such as: data description
and interpretation, inference, pattern recognition, prediction,
classification, compression, and many more.  One way of performing
such tasks is through the use of machine learning methods. For
accessible accounts of machine learning and its use in astronomy, see,
for example, \cite{MacKay_ITILA}, \cite{Ball2010} and
\cite{Way2012}. Moreover, machine learning software easily used for astronomy,
such as the Python-based {\sc astroML} package\footnote{{\tt
http://astroml.github.com/}}, or C-based Fast Artificial Neural Network Library 
(FANN\footnote{\texttt{http://leenissen.dk/fann/wp/}}) have recently started to become
available.

Two major categories of machine learning are: {\em
  supervised learning} and {\em unsupervised learning}.  In supervised
learning, the goal is to infer a function from labeled training data,
which consist of a set of training examples. Each example has both
`properties' and `labels'. The properties are known `input' quantities
whose values are to be used to predict the values of the labels, which
may be considered as `output' quantities. Thus, the function to be
inferred is the mapping from properties to labels. Once learned, this
mapping can be applied to datasets for which the values of the labels
are not known. Supervised learning is usually further subdivided into
{\em classification} and {\em regression}. In classification, the
labels take discrete values, whereas in regression the labels are
continuous.  

In astronomy, for example, using multifrequency observations of a
supernova lightcurve (its properties) to determine its type (e.g. Ia,
Ib, II, etc.)  is a classification problem since the label (supernova
type) is discrete (see, e.g., \citealt{karpenka}), whereas using the
observations to determine (say) the energy output of the supernova
explosion is a regression problem, since the label (energy output) is
continuous. Classification can also be used to obtain a distribution
for an output value that would normally be treated as a regression problem.
This is demonstrated by~\cite{Bonnett2013} for measuring redshifts
in CFHTLenS. A particularly important recent application of regression
supervised learning in astrophysics and cosmology (and beyond) is the
acceleration of the statistical analysis of large data sets in the
context of complicated models. In such analyses, one typically
performs many ($\sim 10^{4-6}$) evaluations of the likelihood function
describing the probability of obtaining the data for different sets of
values of the model parameters.  For some problems, in particular in
cosmology, each such function evaluation can take up to tens of
seconds, making the analysis very computationally expensive. By
performing regression supervised learning to infer and then replace
the mapping between model parameters and likelihood value, once can
reduce the computation required for each likelihood evaluation by
several orders of magnitude, thereby vastly accelerating the analysis
(see, e.g., \citealt{pico, cosmonet1, cosmonet2}).

In unsupervised learning, the data have no labels. More precisely, the
quantities (often termed `observations') associated with each data
item are not divided into properties (inputs) and labels (outputs).
This lack of a `causal structure', where the inputs are assumed to be
at the beginning and outputs at the end of a causal chain, is the key
difference from supervised learning. Instead, all the observations are
considered to be at the end of a causal chain, which is assumed to
begin with some set of `latent' (or hidden) variables. The aim of
unsupervised learning is to infer the number and/or nature of these
latent variables (which may be discrete or continuous) by finding
similarities between the data items. This then enables one to
summarize and explain key features of the dataset.  The most common
tasks in unsupervised learning include {\em density estimation}, {\em
  clustering} and {\em dimensionality reduction}. Indeed, in some
cases, dimensionality reduction can be used as a pre-processing step
to supervised learning, since classification and regression can
sometimes be performed in the reduced space more accurately than in
the original space.

As an astronomical example of unsupervised learning one might wish to
use multifrequency observations of the lightcurves of a set of
supernovae to determine how many different types of supernovae are
contained in the set (a clustering task). Alternatively, if the data
set also includes the type of each supernova (determined using
spectroscopic observations), one might wish to determine which
properties, or combination of properties, in the lightcurves are most
important for determining their type photometrically (a dimensionality
reduction task). This reduced set of property combinations could then
be used instead of the original lightcurve data to perform the
supernovae classification or regression analyses mentioned above.

An intuitive and well-established approach to machine learning, both
supervised and unsupervised, is based on the use of artificial neural
networks (NNs), which are loosely inspired by the structure and
functional aspects of a brain. They consist of a group of
interconnected nodes, each of which processes information that it
receives and then passes this product on to other nodes via weighted
connections. In this way, NNs constitute a non-linear statistical data
modeling tool, which may be used to represent complex relationships
between a set of inputs and outputs, to find patterns in data, or to
capture the statistical structure in an unknown joint probability
distribution between observed variables.  In general, the structure of a
NN can be arbitrary, but many machine learning applications can be
performed using only feed-forward NNs. For such networks the structure is
directed: an input layer of nodes passes information to an output
layer via zero, one, or many `hidden' layers in between.  Such a
network is able to `learn' a mapping between inputs and outputs, given a
set of training data, and can then make predictions of the outputs for
new input data. Moreover, a universal approximation theorem assures us
that we can accurately and precisely approximate the mapping with a NN
of a given form. A useful introduction to NNs can be found
in~\cite{MacKay_ITILA}. 

In astronomy, feed-forward NNs have been applied to various
machine learning problems for over 20 years (see, e.g.,
\citealt{Andreon1999,Andreon2000,LTA2001,Tagliaferri2003a,Tagliaferri2003b, Way2012}). 
Nonetheless, their more widespread use in astronomy
has been limited by the difficulty associated with standard
techniques, such as backpropagation, in training networks having many
nodes and/or numerous hidden layers (i.e. `large' and/or `deep'
networks), which are often necessary to model the complicated mappings
between the numerous inputs and outputs in modern astronomical
applications.

In this paper, we therefore present the first public release of {\sc
  SkyNet}: an efficient and robust neural network training tool that
is able to train large and/or deep feed-forward
networks.\footnote{\SN\,\, may also be used to train recurrent neural
  networks (see, e.g., \citealt{Mandic2001}), but its
  application to such networks will be discussed in a future work.}
{\sc SkyNet} is able to achieve this by using a combination of the
`pre-training' method of \cite{HintonFastRBMmethod} to obtain a set of
network weights close to a good optimum of the training
objective function, followed by further optimisation of the weights
using a regularised variant of Newton's method based on
that developed for the {\sc MemSys} software package
\citep{MemSys}. In particular, second-order derivative information is
used to improve convergence, but without the need to evaluate or store
the full Hessian matrix, by using a fast approximate method to
calculate Hessian-vector products \citep{Schraudolph,
  HessianFree}. {\sc SkyNet} is implemented in the standard ANSI C
programming language and parallelised using MPI.\footnote{A version of
  {\sc SkyNet} parallelised for GPUs using CUDA is currently in
  development.}

We also note that {\sc SkyNet} has already been combined with {\sc
  MultiNest}~\citep{multimodalNS, multinest, multinest3}, to produce
the Blind Accelerated Multimodal Bayesian Inference ({\sc BAMBI})
package \citep{bambi}, which is a generic and completely automated
tool for greatly accelerating Bayesian inference problems (by up to a
factor of $\sim 10^6$; see, e.g., \citealt{coverage}). {\sc MultiNest}
is a fully-parallelised implementation of nested sampling
\citep{Skilling}, extended to handle multimodal and highly-degenerate
distributions.  In most astrophysical (and particle physics) Bayesian
inference problems, {\sc MultiNest} typically reduces the number of
likelihood evaluations required by an order of magnitude or more,
compared to standard MCMC methods, but {\sc BAMBI} achieves further
substantial gains by speeding up the evaluation of the likelihood
itself by replacing it with a trained regression neural network. {\sc
  BAMBI} proceeds by first using {\sc MultiNest} to obtain a specified
number of new samples from the model parameter space, and then uses
these as input to {\sc SkyNet} to train a network on the likelihood
function. After convergence to the optimal weights, the network's
ability to predict likelihood values to within a specified tolerance
level is tested. If it fails, sampling continues using the original
likelihood until enough new samples have been made for training to be
performed again. Once a network is trained that is sufficiently
accurate, its predictions are used in place of the original likelihood
function for future samples for {\sc MultiNest}. On typical problems
in cosmology, for example, using the network reduces the likelihood
evaluation time from seconds to less than a millisecond, allowing {\sc
  MultiNest} to complete the analysis much more rapidly. As a bonus,
at the end of the analysis the user also obtains a network that is
trained to provide more likelihood evaluations near the peak if needed,
or in subsequent analyses. With the public release of {\sc SkyNet}, we
now also make {\sc BAMBI} publically available.\footnote{\tt
  http://www.mrao.cam.ac.uk/software/bambi/}

The structure of this paper is as follows.  In
Section~\ref{sec:NNstruct} we describe the general structure of
feed-forward NNs, including a particular special case of such
networks, called autoencoders, which may be used for performing
non-linear dimensionality reduction.  In Section~\ref{sec:NNtrain} we
present the procedures used by {\sc SkyNet} to train networks of these
types. {\sc SkyNet} is then applied to some toy machine learning
examples in Section~\ref{sec:NNtoyex}, including a regression task, a
classification task, and a dimensionality reduction task using
autoencoders.  We also apply {\sc SkyNet} to the problem of
classifying images of handwritten digits from the MNIST database,
which is a widely-used benchmarking test of machine learning
algorithms. The application of {\sc SkyNet} to astronomical
machine learning examples is presented in
Section~\ref{sec:NNex_astro}, including: a regression task to
determine the projected ellipticity of a galaxy from blurred and noisy
images of the galaxy and of a field star; a classification task, based
on a simulated gamma-ray burst detection pipeline for the Swift
satellite \citep{Gehrels2004}, to determine if a GRB with given source
parameters will be detected; and a dimensionality reduction task using
autoencoders to compress and denoise galaxy images.  Finally, we present our
conclusions in Section~\ref{sec:NNdiscuss}.

\section{Network structure}\label{sec:NNstruct}

\subsection{Feed-forward neural networks}

\begin{figure}
\begin{center}
\includegraphics[width=0.55\columnwidth]{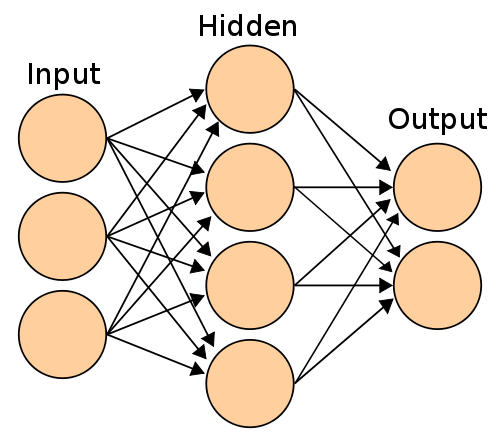}
\caption{A 3-layer neural network with 3 inputs, 4 hidden nodes, and 2 outputs. Image courtesy of Wikimedia Commons.}
\label{fig:neuralnet}
\end{center}
\end{figure}

A multilayer perceptron feed-forward neural network is the simplest
type of network and consists of ordered layers of perceptron nodes
that pass scalar values from one layer to the next. The perceptron is
the simplest kind of node, and maps an input vector $\bmath{x} \in
\Re^n$ to a scalar output $f(\bmath{x};\bmath{w},\theta)$ via
\begin{equation}
\label{eq:perceptron}
f(\bmath{x};\bmath{w},\theta) = \theta + \sum_{i=1}^n {w_i x_i},
\end{equation}
where $\bmath{w}=\{w_i\}$ and $\theta$ are the parameters of the
perceptron, called the `weights' and `bias', respectively. For a
3-layer NN, which consists of an input layer, a hidden layer, and an
output layer, as shown in Fig.~\ref{fig:neuralnet}, the outputs of
the nodes in the hidden and output layers are given by the following
equations:
\begin{eqnarray}
\label{eq:hiddenformula}
\textrm{hidden layer:} \; h_j = g^{(1)}(f^{(1)}_j); \; f^{(1)}_j = \theta^{(1)}_j + \sum_l {w^{(1)}_{jl}x_l} ,\\
\label{eq:outputformula}
\textrm{output layer:} \; y_i = g^{(2)}(f^{(2)}_i); \; f^{(2)}_i = \theta^{(2)}_i + \sum_j {w^{(2)}_{ij}h_j} ,
\end{eqnarray}
where $l$ runs over input nodes, $j$ runs over hidden nodes, and $i$
runs over output nodes. The functions $g^{(1)}$ and $g^{(2)}$ are
called activation functions and must be smooth and monotonic
for our purposes. We use $g^{(1)}(x)=\slfrac{1}{(1+e^{-x})}=\sig(x)$
(sigmoid) and $g^{(2)}(x)=x$; the non-linearity of $g^{(1)}$ is
essential to allowing the network to model non-linear functions.  To
expand the NN to include more hidden layers, we iterate
\eqref{eq:hiddenformula} for each connection from one hidden layer to
the next, each time using the same activation function, $g^{(1)}$. The
final hidden layer will connect to the output layer using the relation
\eqref{eq:outputformula}.

The weights and biases are the values we wish to determine in our
training (described in Section~\ref{sec:NNtrain}). As they vary, a
huge range of non-linear mappings from inputs to outputs is
possible. In fact, a universal approximation theorem~\citep{UnivApprox2,UnivApprox}
states that a NN with three or more layers can approximate any
continuous function to some given accuracy, as long as the activation
function is locally bounded, piecewise continuous, and not a
polynomial (hence our use of sigmoid $g^{(1)}$, although other
functions would work just as well, such as $\tanh$). By increasing
the number of hidden nodes, one can achieve more accuracy at the risk
of overfitting to our training data.

Other activation functions have also been proposed, such as the rectified linear function wherein
$g(x) = \textrm{max}\{0,x\}$ or the `softsign' function where $g(x) = \slfrac{x}{\left(1+\lvert x \rvert\right)}$.
It has been argued that the former removes the need for pre-training (as described in Section~\ref{sec:NNpretrain})~\citep{Glorot2011ReLU}
and serves as a better model of biological neurons.
The `softsign' is similar to $\tanh$, but with slower approach to the asymptotes of $\pm1$
(quadratic rather than exponential)~\citep{Bergstra2009,GlorotBengio2010}.

\subsection{Autoencoders}\label{sec:autoenc}

\begin{figure}
\begin{center}
\includegraphics[width=\columnwidth]{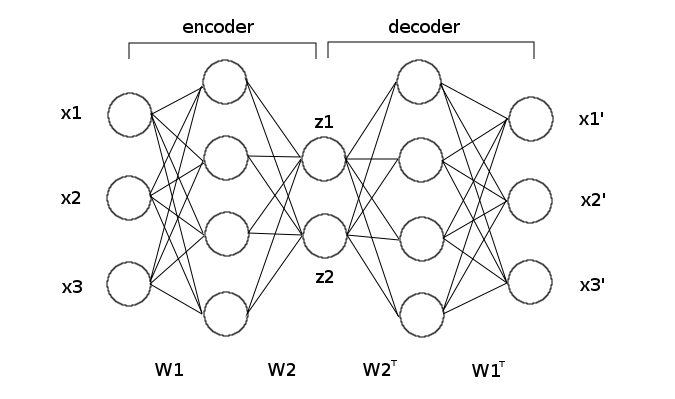}
\caption{Schematic diagram of an autoencoder. The three input values are
  encoded to two feature variables. Pre-training (described in
  Section~\ref{sec:NNpretrain}) defines the weight matrices $W_1$ and $W_2$.}
\label{fig:autoencoder}
\end{center}
\end{figure}

Autoencoders are a specific type of feed-forward neural network
containing one or more hidden layers, where
the inputs are mapped to themselves,
i.e. the network is trained to approximate the identity operation;
when more than one hidden layer is used this is typically referred to as a 
`deep' autoencoder.
Such networks typically contain several hidden layers and are
symmetric about a central layer containing fewer nodes than there are
inputs (or outputs).  A basic diagram of an autoencoder is shown in
Fig.~\ref{fig:autoencoder}, in which the three inputs $(x_1,x_2,x_3)$ are
mapped to themselves via three symmetrically-arranged hidden layers,
with two nodes in the central layer.

An autoencoder can thus be considered as two half-networks, with one
part mapping the inputs to the central layer and the second part
mapping the central layer values to the outputs (which approximate as
closely as possible the original inputs). These two parts are called
the `encoder' and `decoder', respectively, and map either to or from a
reduced set of `feature variables' embodied in the nodes of the
central layer (denoted by $z_1$ and $z_2$ in
Fig.~\ref{fig:autoencoder}). These variables are, in general,
non-linear functions of the original input variables.  One can
determine this dependence for each feature variable in turn simply by
decoding $(z_1,0,0,\ldots,0)$, $(0,z_2,0,\ldots,0)$, and so on, as the
corresponding $z_i$ value is varied; in this way, for each feature
variable, one obtains a curve in the original data space. Conversely,
the collection of feature values $(z_1,z_2,\ldots,z_M)$ in the central
layer might reasonably be termed the feature vector of the input
data. Autoencoders therefore provide a very intuitive approach to
non-linear dimensionality reduction and constitute a natural
generalisation of linear methods such as principal component analysis
(PCA) and independent component analysis (ICA), which are widely used
in astronomy. Indeed, an antoencoder with a single hidden layer and
linear activation functions may be shown to be identical to PCA
\citep{sanger}. This topic is explored further in
Section~\ref{sec:NNtoy_AE}. It is worth noting that encoding from
input data to feature variables can also be useful in performing
clustering tasks; this is illustrated in Section~\ref{sec:MNIST}.

Autoencoders are, however, notoriously difficult to train, since the
objective function contains a broad local maximum where each output
is the average value of the inputs~\citep{Erhan2010}. Nonetheless, this difficulty can
be overcome by the use of pre-training methods, as discussed in
Section~\ref{sec:NNpretrain}.

\subsection{Choosing the number of hidden layers and nodes}\label{sec:nhid}

An important choice when training a NN is the number of nodes in its
hidden layers. The optimal number and organisation into one or more
layers has a complicated dependence on the number of training data
points, the number of inputs and outputs, and the complexity of the
function to be trained. Choosing too few nodes will mean that the NN
is unable to learn the relationship to the highest possible accuracy;
choosing too many will increase the risk of overfitting to the
training data and will also slow down the training process. Using
empirical evidence~\citep{murtagh} and theoretical considerations~\citep{GevaSitte},
it has been suggested that the optimal architecture
for approximating a continuous function is one hidden layer containing
$2N+1$ nodes, where $N$ is the number of input nodes. \cite{AstroNN}
also find empirical support for this suggestion. Such a choice allows
the network to model the form of the mapping function without
unnecessary work.

In practice, it can be better to over-estimate (slightly) the number
of hidden nodes required. As described in Section~\ref{sec:NNtrain},
{\sc SkyNet} performs basic checks to prevent over-fitting, and the
additional training time associated with having more hidden nodes is
not a large penalty if an optimal network can be obtained in an early
attempt.  In any case, given a particular problem, the optimal network
structure, both in terms of the number of hidden nodes and how they are
distributed into layers, can be determined by comparing the
correlation and error squared of different trained NNs; this is
illustrated in Section~\ref{sec:NNtoyex}.

\section{Network training}\label{sec:NNtrain}

In training a NN, we wish to find the optimal set of network weights
and biases that maximise the accuracy of the predicted
outputs. However, we must be careful to avoid overfitting to our
training data, which may lead to inaccurate predictions from inputs
on which the network has not been trained.

The set of training data inputs and outputs (or `targets'),
$\mathcal{D} = \{\bmath{x}^{(k)},\bmath{t}^{(k)}\}$, is provided by
the user (where $k$ counts training items). Approximately $75$ per
cent should be used for actual NN training and the remainder retained
as a validation set that will be used to determine convergence and to
avoid overfitting. This ratio of 3:1 gives plenty of information for
training but still leaves a representative subset of the data for
checks to be made.

\subsection{Data whitening}\label{sec:NNtrain_whiten}

It is prudent to `whiten' the data before training a
network. Whitening normalises the input and/or output values, so that
it easier to train a network starting from initial weights that are
small and centred on zero. The network weights in the first and last
layers can then be `unwhitened' after training so that the network
will be able to perform the mapping from original inputs to outputs.

Standard whitening transforms each input to a standard 
distribution by subtracting the mean and dividing by the standard
deviation over all elements in the training data, such that
\begin{subequations}
\label{eq:whitening1}
\begin{align}
\label{eq:whitening1a}
\tilde{x}_l^{(k)} &= \frac{x_l^{(k)} - \overline{x}_l}{\sigma_{l}}, \\
\label{eq:whitening1b}
\overline{x}_l &= \frac{1}{K} \sum_{k=1}^K {x_l^{(k)}}, \\
\label{eq:whitening1c}
\sigma_l^2 &= \frac{1}{K-1} \sum_{k=1}^K {(x_l^{(k)} - \overline{x}_l)^2}.
\end{align}
\end{subequations}
An alternative whitening transform is also commonly used, wherein all
values are scaled and shifted into the interval $[0,1]$, such that
\begin{equation}
\label{eq:whitening2}
\tilde{x}_l^{(k)} = \frac{x_l^{(k)} - \textrm{min}_k(x_l^{(k)})}{\textrm{max}_k(x_l^{(k)}) - \textrm{min}_k(x_l^{(k)})}.
\end{equation}

One of these transforms may be chosen by the user if they wish to
whiten the inputs of the training data. The whitening is normally
performed separately on each input, but can be calculated across all
inputs if they are related. The mean, standard deviation, minimum, or
maximum would then be computed over all inputs for all training data
items. The chosen whitening transform is also used for whitening the
outputs. Since both transforms consist of subtracting an offset and
multiplying by a scale factor, they can easily be performed and
reversed. To unwhiten network weights the inverse transform is
applied, with the offset and scale determined by the source input node
or target output node. Outputs for a classification network are not
whitened since they are already just probabilities (see below).

\subsection{Network objective function}\label{sec:NNtrain_overview}

Let us denote the network weights and biases collectively by the
network parameter vector $\bmath{a}$. {\sc SkyNet} considers the
parameters $\bmath{a}$ to be random variables with a posterior
distribution given by
\begin{equation}
\mathcal{P}(\bmath{a};\alpha,\bm{\sigma}) \propto 
\mathcal{L}(\bmath{a};\bm{\sigma}) \times \mathcal{S}(\bmath{a};\alpha),
\label{eq:netpost}
\end{equation}
where $\mathcal{L}(\bmath{a};\bm{\sigma})$ is the likelihood, which also depends on a set of hyperparameters
$\bm{\sigma}$ that describe the standard deviation of the outputs (see
below).  The likelihood encodes how well the NN, characterised by a
given set of parameters $\bmath{a}$, is able to reproduce the known
training data outputs. This is modulated by the prior
$\mathcal{S}(\bmath{a};\alpha)$, which is assumed to have the
(logarithmic) form
\begin{equation}
\label{eq:netprior}
\log\mathcal{S}(\bmath{a};\alpha) = -\frac{\alpha}{2} \sum_i {a_i^2},
\end{equation}
where $\alpha$ is a hyperparameter that plays the role of
a regularisation parameter during optimisation since it determines
relative importance of the prior and the likelihood. This prior 
can also be seen as an $\ell^2$-norm penalty. The form of the
likelihood depends on the type of network being trained.

\subsubsection{Regression likelihood}

For regression problems, {\sc SkyNet} assumes a log-likelihood
function for the network parameters $\bmath{a}$ given by the 
standard $\chi^2$ misfit function
\begin{align}
\label{eq:chisquared}
\log\mathcal{L}(\bmath{a};\bm{\sigma}) =& -\frac{K\log(2\pi)}{2}-\sum_{i=1}^N {\log(\sigma_i)} \notag \\
&-\frac{1}{2} \sum_{k=1}^K \sum_{i=1}^N {\left[\frac{t_i^{(k)} - y_i(\bmath{x}^{(k)};\bmath{a})}{\sigma_i}\right]^2},
\end{align}
where $N$ is the number of outputs, $K$ is the number of training data
examples and $\bmath{y}(\bmath{x}^{(k)};\bmath{a})$ is the NN's
predicted output vector for the input vector $\bmath{x}^{(k)}$ and
network parameters $\bmath{a}$. The hyperparameters
$\bm{\sigma}=\{\sigma_i\}$ describe the standard deviation (error
size) of each of the outputs.

\subsubsection{Classification likelihood}
For classification problems, {\sc SkyNet} again uses continuous
outputs (rather than discrete ones), which are interpreted as the
probabilities that a set of inputs belongs to a particular output
class. This is achieved by applying the {\em softmax} transformation
to the output values, so that they are all non-negative and sum to
unity, namely
\begin{equation}
\label{eq:classnetout}
y_i(\bmath{x}^{(k)};\bmath{a}) \to \frac{\exp[y_i(\bmath{x}^{(k)};\bmath{a})]}{\sum_{j=1}^N \exp[y_j(\bmath{x}^{(k)};\bmath{a})]}.
\end{equation}
The classification likelihood is then given by the {\em cross-entropy} of the 
targets and softmaxed output values~\citep{MacKay_ITILA},
\begin{equation}
\label{eq:classlike}
\log\mathcal{L}(\bmath{a};\bm{\sigma}) = -\sum_{k=1}^K \sum_{i=1}^N t_i^{(k)} \log y_i(\bmath{x}^{(k)};\bmath{a}).
\end{equation}
In this scenario, the true and predicted output values are
both probabilities (which lie in $[0,1]$). For the true outputs, all are
zero except for the correct output which has a value of unity. For
classification networks, the $\bm{\sigma}$ hyper-parameters do not
appear in the log-likelihood.

\subsection{Initialisation and pre-training}\label{sec:NNpretrain}


The training of the NN can be started from some random initial state,
or from a state determined from a `pre-training' procedure discussed
below.  In the former case, the network training begins by setting
random values for the network parameters, sampled from a normal
distribution with zero mean and variance of $0.01$ (this value can be
modified by the user).

\begin{figure}
\begin{center}
\includegraphics[width=0.5\columnwidth]{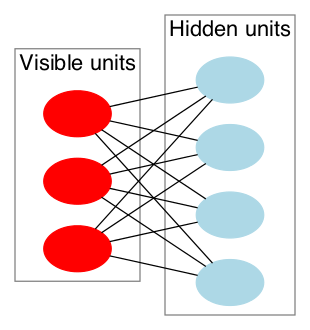}
\caption{Diagram of an RBM with $3$ visible nodes and $4$ hidden nodes. Bias nodes are not shown. Image courtesy Wikimedia Commons.}
\label{fig:RBM}
\end{center}
\end{figure}

In the latter case, {\sc SkyNet} makes use of the pre-training
approach developed by \cite{HintonFastRBMmethod}, which obtains a set
of network weights and biases close to a good solution of the
network objective function.  This method was originally devised with
autoencoders in mind and is based on the model of restricted Boltzmann machines (RBMs).
An RBM is a generative model that can learn a probability distribution over a set of
inputs. It consists of a layer of input nodes and a layer of hidden nodes, as shown
in Figure~\ref{fig:RBM}. In this case, the map from the inputs to the
hidden layer and then back is treated symmetrically and the
weights are adjusted through a number of `epochs', gradually reducing
the reproduction error. To model an autoencoder, RBMs are `stacked', with each
RBM's hidden layer being the input for the next. The initial case is the NN's inputs
to the first hidden layer; this is repeated for the first to second
hidden layer and so on until the central layer is reached. The network
weights can then be `unfolded' by using the transpose for the
symmetric connections in the decoding half to provide a decent
starting point for the full training to begin. This is shown in
Fig.~\ref{fig:autoencoder}, where the $W_1$ and $W_2$ weights matrices
are defined by pre-training.

The training is then performed using \textit{contrastive divergence}~\citep{contrastdiv}.
This procedure can be summarised in the following steps, where sampling indicates setting the
value of the node to $1$ with the probability calculated and $0$ otherwise.
\begin{enumerate}
\item Take a training sample ${\bf x}$ and compute the probabilities of the hidden nodes (their values using a sigmoid activation function) and sample a hidden vector ${\bf h}$ from this distribution.
\item Let $g_{+} = {\bf x} \otimes {\bf h}$, where $\otimes$ is used to indicate the outer product.
\item Using ${\bf h}$, compute the probabilties of the visible nodes and sample ${\bf x}^\prime$ from this distribution. Resample the hidden vector from this to obtain ${\bf h}^\prime$.
\item Let $g_{-} = {\bf x}^\prime \otimes {\bf h}^\prime$.
\item $w_{i,j} \mapsto w_{i,j} + r (g_{+} - g_{-})$ for some learning rate $0 < r \leq 1$.
\end{enumerate}

More details can be found
in~\cite{HintonFastRBMmethod} and~\cite{Hinton_AE_Science} has useful
diagrams and explanations.

This pre-training approach can also be used for more general
feed-forward networks. All layers of weights, except for the final one
that connects the last hidden layer to the outputs, are pre-trained as
if they were the first half of a symmetric autoencoder. However, the
network weights are not unfolded; instead the final layer of weights
is initialised randomly as would have been done without
pre-training. In this way, the network `learns the inputs' before
mapping to a set of outputs. This has been shown to greatly reduce the
training time on multiple problems by~\cite{GlorotBengio2010,Erhan2010}.

We note that when an autoencoder is pre-trained, the activation
function to the central hidden layer is made linear and the activation
function from the final hidden layer to the outputs is made
sigmoidal. General feed-forward networks that are pre-trained continue
to use the original activation functions. Both of these are simply
the default settings and the user has the freedom to alter them
to suit their specific problem.

\subsection{Optimisation of the objective function}

Once the initial set of network parameters have been obtained, either
by assigning them randomly or through pre-training, the network is
then trained (further) by iterative optimisation of the objective function.

First, initial values of the hyperparameters $\bm{\sigma}$ (for
regression networks) and $\alpha$ are set.  The values $\bm{\sigma}$
are set by the user and can be set on either the true output
values themselves or on their whitened values (as defined in
Section~\ref{sec:NNtrain_whiten}). The only difference between these
two settings is the magnitude of the error used. The algorithm then
calculates a large initial estimate for $\alpha$,
\begin{equation}
\alpha = \frac{\lvert \nabla \log(\mathcal{L}) \rvert}{\sqrt{M r}},
\label{eq:initalpha}
\end{equation}
where $M$ is the total number of weights and biases (NN parameters)
and $r$ is a rate set by the user ($0 < r \leq 1$, default $r=0.1$)
that defines the size of the `confidence region' for the
gradient. This expression for $\alpha$ sets larger regularisation
(or `damping') when the magnitude of the gradient of the
likelihood is larger. This relates the amount of `smoothing'
required to the steepness of the function being smoothed. The rate
factor in the denominator allows us to increase the damping for
smaller confidence regions on the value of the gradient. This results
in smaller, more conservative steps that are more likely to result in
an increase in the function value but results in more steps being
required to reach the optimal weights.

NN training then proceeds using an adapted form of a truncated Newton
(or `Hessian-free') optimisation algorithm as described below, to
calculate the step, $\delta\bmath{a}$, that should be taken at each
iteration.  Following each such step, adjustments to $\alpha$ and
$\bm{\sigma}$ may be made before another step is calculated.
First, $\bm{\sigma}$ can be updated by multiplying it
by a value $c$ such that $c^2 = \slfrac{-2(\log\mathcal{L}+\log\mathcal{S})}{M}$.
This serves to assure that at convergence, the $\chi^2$ value equals the number of
unconstrained data points of the problem. Similarly, $\alpha$ is then updated such that
the probability $\Pr(\mathcal{D}\vert\alpha)$ is maximised for the current set of NN
parameters $\bmath{a}$. These procedures are described in detail by~\citet[Sec. 2.3 \& 2.6]{MemSys}
and~\citet[Sec. 3.6 \& Appendix B]{Hobson98}.

To obtain the step $\delta\bmath{a}$ at each iteration, we first note
that one may approximate a general function $f$ up to second-order in
its Taylor expansion by
\begin{equation}
f(\bmath{a}+\delta\bmath{a}) \approx f(\bmath{a}) + \bmath{g}^{\textrm{T}} \delta\bmath{a} + {\textstyle\frac{1}{2}} (\delta \bmath{a})^{\textrm{T}} \bsf{B} 
\,\delta \bmath{a},
\label{eq:newton}
\end{equation}
where $\bmath{g}=\nabla f(\bmath{a})$ is the gradient and
$\bsf{B}=\nabla\nabla f(\bmath{a})$ is the Hessian matrix of second
derivatives, both evaluated at $\bmath{a}$.  For our purposes, the
function $f$ is the log-posterior distribution of the NN parameters
and hence~\eqref{eq:newton} represents a Gaussian approximation to the
posterior. The Hessian of the log-posterior is the regularised
(`damped') Hessian of the log-likelihood function, where the prior,
whose magnitude is set by $\alpha$, provides the regularisation. If we
define the Hessian matrix of the log-likelihood as $\bsf{H}$, then
$\bsf{B}=\bsf{H}+\alpha \bsf{I}$, where $\bsf{I}$ is the identity
matrix.  The regularisation parameter $\alpha$ can be interpreted as
controlling the level of `conservatism' in the Gaussian approximation
to the posterior. In particular, regularisation helps prevent the
optimisation becoming trapped in small local maxima by smoothing out
the function being explored. It also aids in reducing the region of
confidence for the gradient information which will make it less likely
that a step results in a worse set of parameters.

Ideally, we seek a step $\delta\bmath{a}$, such that $\nabla
f(\bmath{a}+\delta \bmath{a}) = 0$.  Using the approximation
\eqref{eq:newton}, one thus requires
\begin{equation}
\bsf{B}\, \delta \bmath{a} = -\bmath{g}.
\label{eq:newtongrad}
\end{equation}
In the standard Newton's method of optimisation one simply solves
this equation directly for $\delta \bmath{a}$ to obtain
\begin{equation}
\delta \bmath{a} = -\bsf{B}^{-1} \bmath{g}.
\label{eq:newtonsoln}
\end{equation}
In principle, iterating this stepping procedure will eventually bring
us to a local maximum of $f$. Moreover, Newton's method has the
important property of being scale-invariant, namely its behaviour is
unchanged under any linear rescaling of the parameters. Methods
without this property often have problems optimising poorly scaled
parameters.

There are, however, some major practical difficulties with the
standard Newton's method. First, the Hessian $\bsf{H}$ of the
log-likelihood is not guaranteed to be positive semi-definite. Thus,
even after the addition of the damping term $\alpha\bsf{I}$ derived
from the log-prior, the full Hessian $\bsf{B}$ of the log-posterior
may also not be invertible. Second, even if $\bsf{B}$ is invertible, the
inversion is prohibitively expensive if the number of parameters is
large, as is the case even for modestly-sized neural networks.

To address the first issue, we replace the Hessian $\bsf{H}$ with a
form of Gauss--Newton approximation $\bsf{G}$, which is guaranteed to
be positive semi-definite and can be defined both for the regression
likelihood \eqref{eq:chisquared} and the classification likelihood
\eqref{eq:classlike}, respectively \citep{Schraudolph}. In particular,
the approximation used differs from the classical Gauss--Newton matrix
in that it retains some second derivative information.  Second, to
avoid the prohibitive expense of calculating the inverse in
\eqref{eq:newtonsoln}, we instead solve \eqref{eq:newtongrad} (with
$\bsf{H}$ replaced by $\bsf{G}$ in $\bsf{B}$) for $\delta\bmath{a}$
iteratively using a conjugate-gradient algorithm, which requires only
matrix-vector products $\bsf{B}\bmath{v}$ for some vector $\bmath{v}$.

One can avoid even the computational burden of calculating and storing
the Hessian $\bsf{B}$. In principle, products of the form
$\bsf{B}\bmath{v}$ can be easily computed using finite differences at
the cost of a single extra gradient evaluation using the identity
\begin{equation}
\bsf{B}\bmath{v} = \lim_{r\to 0}\frac{\nabla f(\bmath{a}+r\bmath{v}) - 
\nabla f(\bmath{a})}{r}.
\label{eq:Hessianvectorprod1}
\end{equation}
This approach is, however, subject to numerical problems.
Therefore, we instead calculate $\bsf{B}\bmath{v}$ products using a
stable and efficient procedure
applicable to NNs \citep{Pearlmutter, Schraudolph}.  This involves an
additional forward and backward pass through the network beyond
the initial ones required for a gradient calculation.

The combination of all the above methods makes practical the use of
second-order derivative information even for large networks and
significantly improves the rate of convergence of NN training over
standard backpropagation methods.

It has been noted that this method for quasi-Newton second-order descent is
equivalent to the first-order `natural gradient' by~\cite{Pascanu2013}.

\subsection{Convergence}\label{sec:NNtrain_converge}

Following each iteration of the optmisation algorithm, the posterior,
likelihood, correlation, and error squared values are calculated both
for the training data and for the validation data (which were not used
in calculating the steps in the optimisation). The correlation of the
network outputs is defined for each output $i$ as
\begin{equation}
\mbox{Corr}_i(\bmath{a}) = \frac{\sum_{k=1}^K {(t_i^{(k)} - \overline{t}_i)(y_i - \overline{y}_i)}}{\sqrt{\sum_{k=1}^K {(t_i^{(k)} - \overline{t}_i)^2}\sum_{k=1}^K {(y_i^{(k)} - \overline{y}_i)^2}}},
\label{eq:corr_defn}
\end{equation}
where $\overline{t}_i$ and $\overline{y}_i$ are the means of these
output variables over all the training data; the functional
dependencies of $y_i^{(k)}$ have been dropped for brevity. The
correlation provides a relative measure of how well the predicted
outputs match the true ones.  In practice, the correlations from each
output can be averaged together to give an average correlation for the
network's predictions. The average error-squared of the network
outputs is defined by
\begin{equation}
\mbox{ErSq}({\bf a}) = \frac{1}{NK} \sum_{k=1}^K \sum_{i=1}^N {\left[t_i^{(k)} - y_i({\bf x}^{(k)};{\bf a})\right]^2},
\label{eq:errsqr_defn}
\end{equation}
and is complementary to their correlation, since it is an
absolute measure of accuracy.

As one might expect, as the optimisation proceeds, there is a steady
increase in the values of the posterior, likelihood, correlation, and
negative of the error squared, evaluated both for the training and
validation data. Eventually, however, the algorithm will begin to
overfit, resulting in the continued increase of these quantities when
evaluated on the training data, but a decrease in them when evaluated
on the validation data. This divergence in behaviour is taken as
indicating that the algorithm has converged and the optimisation in
terminated. The user may choose which of the four quantities listed
above is used to determine convergence, although the default is to use
the error squared, since it does not include the hyperparameters
$\bsigma$ and $\alpha$ in its calculation and is less prone to problems
with zeros than the correlation. 

\subsection{Optimising network structure}

We note that the correlation and the error-squared functions discussed
above also provide quantitative measures with which to compare the
performance of different network architectures, both in terms of the
number of hidden nodes and how they are distributed into layers.  As
network size and complexity is increased, a point will be reached at
which minimal or no further gains may be achieved in increasing
correlation or reducing error-squared. Therefore, any network
architecture that can achieve this peak performance is equally
well-suited. In practice, we will wish to find the smallest or
simplest network that does so as this minimizes the risk of
overfitting and the time required for training.


\subsection{Estimating the error on network outputs}

After training a network, in particular a regression network, one may
want to calculate the accuracy of the network's predicted outputs. A
computationally cheap method of estimating this was suggested
by~\cite{NNprederror}, whereby one adds Gaussian noise to
the true outputs of the training data and trains a new network on this
noisy data. After performing many realisations, the networks'
predictions will average to the predictions in the absence of the
added noise. Moreover, the standard deviation of their predictions
will provide a good estimate of the accuracy of the original network's
predictions. Since one can train the new networks using the original
trained network as a starting point, the re-training on noisy data is
very fast. Additionally, evaluating the ensemble of predictions to
measure the accuracy is not very computationally intensive as network
evaluations are simple to perform and can be done in less than a
millisecond. Explicitly, the steps of this method are:
\begin{enumerate}
\item Start with the converged network with parameters
  $\bmath{a}^{\ast}$, trained on the original data set $D^{\ast} =
  \{\bmath{x}^{(k)},\bmath{t}^{(k)}\}$. Estimate the noise on the
  residuals using $\sigma^2 = \sum_k
  [t^{(k)}-y(\bmath{x}^{(k)},\bmath{a}^{\ast})]^2/K$.
\item Define a new data set $D^1$ by adding Gaussian noise of zero
  mean and variance $\sigma^2$ to the outputs (targets) in $D^{\ast}$.
\item Train a NN on $D^1$ using the parameters $\bmath{a}^{\ast}$ as a starting
  point. Training should converge rapidly as the new data set is only
  slightly different from the original. Denote the new network parameters by 
$\bmath{a}^1$.
\item Repeat steps (ii) and (iii) multiple times to obtain an ensemble of
  networks with parameters  $\bmath{a}^j$.
\item Use each of the networks $\bmath{a}^j$ to make a prediction for
  a given set of inputs. The accuracy of the original network's
  outputs can be estimated as the standard deviation of the outputs of
  these networks.
\end{enumerate}

In addition to these steps, {\sc SkyNet} includes the option for the
user to add random Gaussian offsets to the parameters
$\bmath{a}^{\ast}$ before training is performed on the new data set
(step iii). This offset will aid the training in moving the
optimisation from a potential local maximum in the posterior
distribution of the network parameters, but the size of the offset
must be chosen for each problem; for this, we recommend using a value
$s \lesssim \slfrac{1}{\alpha}$. We thus add noise to both the
training data and the saved network parameters before training a new
network whose posterior maximum will be near to, but not exactly the
same as, the original network's.

This method may be compared with that described in~\cite{bambi} for
determining the accuracy of the NN predictions for the likelihood used
in {\sc BAMBI}. Although the method described here requires the
overhead time of training additional networks, this is small compared
the speed gains possible. Indeed, the new method's accuracy
computations require less than a millisecond, as opposed to tenths of
a second for the method used previously. Consequently, the faster
method described here is now incorporated into our new public release
version of {\sc BAMBI}, leading to around two orders of magnitude
increase in speed over that reported in~\cite{bambi}.

\section{Applications:  Toy Examples}\label{sec:NNtoyex}

\subsection{Regression}\label{sec:NNtoy_sinc}

As our first toy example, we consider a simple regression problem. 
We generate $200$ points randomly in the range $x \in [-5\pi,5\pi]$, for
which we evaluate the ramped sinc function,
\begin{equation}
\label{eq:sincform}
y(x) = \frac{\sin(x)}{x} + 0.04x,
\end{equation}
and then add Gaussian noise with zero mean and a standard deviation of
$0.05$. The addition of noise makes the regression problem more
difficult and prevents any exact solution being possible.

\begin{figure}
\begin{center}
\includegraphics[width=0.9\columnwidth,height=5.7cm]{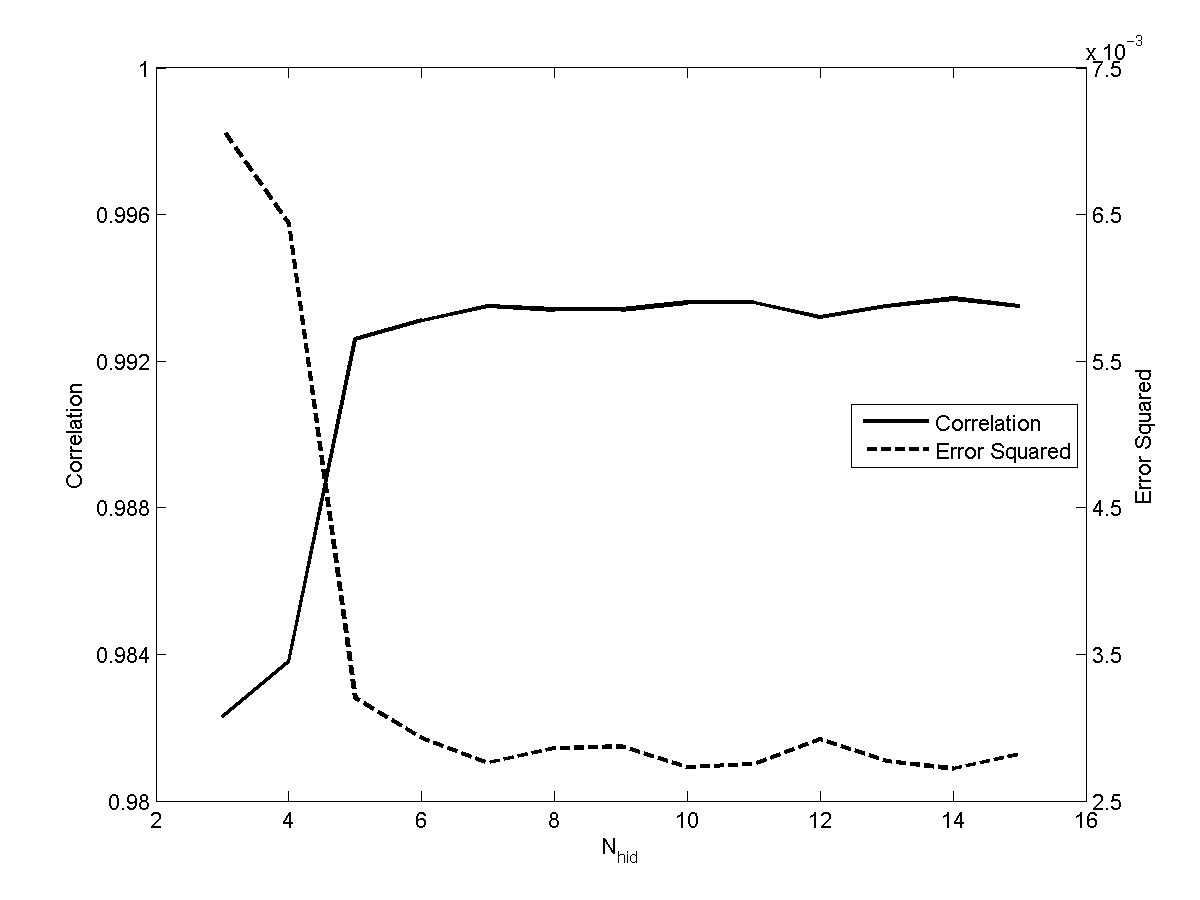}
\caption{The correlation and error-squared values as a function of the
  number of hidden nodes $N$ obtained from converged NNs with
  architecture $1+N+1$ for the ramped sinc function regression
  problem.}
\label{fig:sincevidence}
\end{center}
\end{figure}

\begin{figure}
\begin{center}
\hspace*{-0.6cm}\includegraphics[width=1.15\columnwidth]{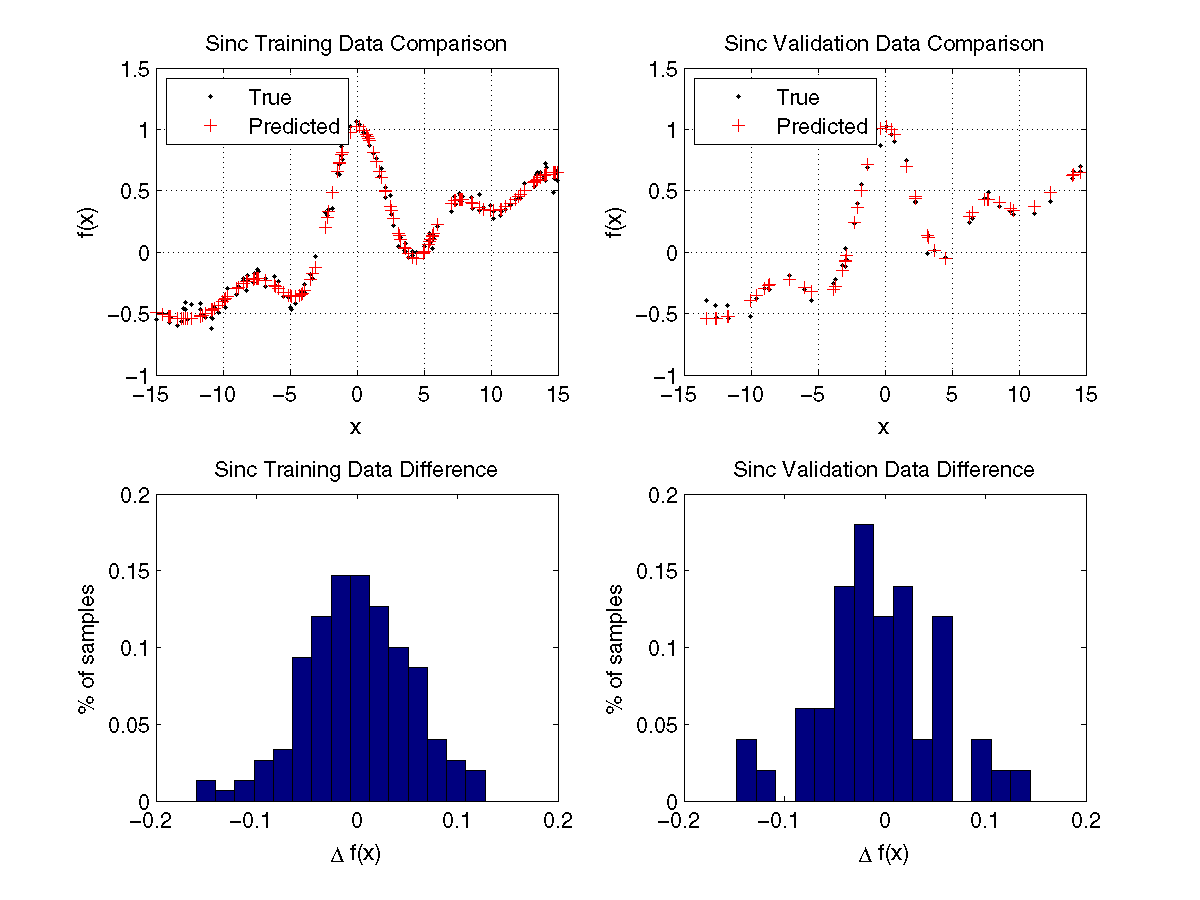}
\caption{Comparisons of the true and predicted values obtained from
  the converged NN with architecture $1+7+1$ on the training data
  (left) and validation data (right) for the ramped sinc function
  regression problem.}
\label{fig:sincplot}
\end{center}
\end{figure}

To perform the regression, the $200$ data items $(x,y)$ are divided
randomly into $150$ items for training and $50$ for validation. For
this simple problem, we use a network with a single hidden layer
containing $N$ nodes (we denote the full network by $1+N+1$), and we
whiten the input and output data using \eqref{eq:whitening1}. The
network was not pre-trained. The optimal value for $N$ is determined
by comparing the correlation and error-squared for networks with
different numbers of hidden nodes.  These results are presented in
Fig.~\ref{fig:sincevidence}, which shows that the correlation
increases and the error-squared decreases until we reach $N\approx
6-7$ hidden nodes, after which both measures level off.
Thus, adding additional nodes beyond this number does not improve
the accuracy of the network. For the network with $N=7$ hidden nodes, 
we obtain a correlation of greater than $99.3$ per cent; a comparison
of the true and predicted outputs in this case is shown in
Figure~\ref{fig:sincplot}.

\subsection{Classification}\label{sec:NNtoy_classify}

We now consider a toy classification problem based on
the three-way classification data set created by Radford 
Neal\footnote{\texttt{http://www.cs.toronto.edu/$\sim$radford/ \linebreak
    fbm.2004-11-10.doc/Ex-netgp-c.html}} for testing his own
algorithms for NN training. In this data set, each of four variables
$x_1$, $x_2$, $x_3$, and $x_4$ is drawn $1000$ times from the standard
uniform distribution $\mathcal{U}[0,1]$. If the two-dimensional
Euclidean distance between $(x_1,x_2)$ and $(0.4,0.5)$ is less than
$0.35$, the point is placed in class $0$; otherwise, if $0.8x_1+1.8x_2
< 0.6$, the class was set to $1$; and if neither of these conditions
is true, the class was set to $2$. Note that the values of $x_3$ and
$x_4$ play no part in the classification. Gaussian noise with zero
mean and standard deviation $0.1$ is then added to the input values.

Approximately $75$ percent of the data was used for training and the
remaining $25$ per cent for validation. We again use a network with a
single hidden layer containing $N$ nodes, and we whiten the input and
output data using \eqref{eq:whitening1}. The network was not
pre-trained. The full network
thus has the architecture $4 + N + 3$, where the three output nodes
give the probabilities (which sum to unity) that the input data
belong to class 0, 1, or 2, respectively. The final class assigned is
that having the largest probability.

\begin{figure}
\begin{center}
\includegraphics[width=0.9\columnwidth,height=5.7cm]{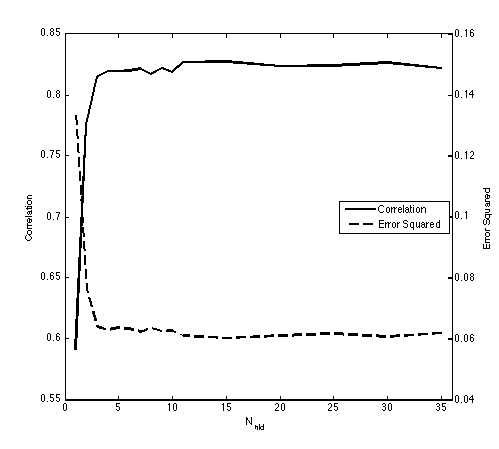}
\caption{The correlation and error squared of the converged NNs for the classification problem as a function of nodes in the single hidden layer.}
\label{fig:classprobstruct}
\end{center}
\end{figure}

\begin{figure}
\begin{center}
\includegraphics[width=\columnwidth,height=0.9\columnwidth]{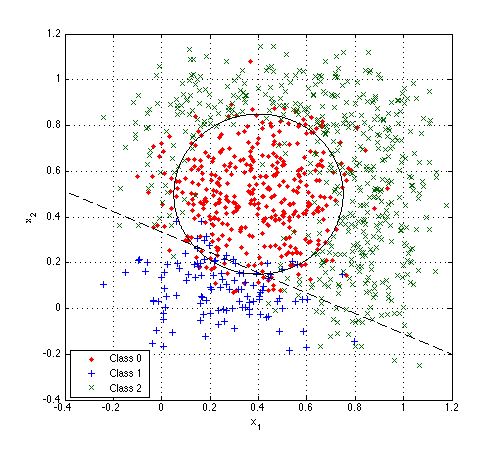}
\caption{The full Neal data set (training and validation), showing the
  NN classifications (colour-coded) and the true criteria (solid and
  dashed lines) determining the classes of the data in the absence of
  noise.}
\label{fig:cdataplot}
\end{center}
\end{figure}

The optimal value for $N$ is again determined by comparing the
correlation and error-squared for networks with different numbers of
hidden nodes.  These results are shown in
Fig.~\ref{fig:classprobstruct}, from which we
see that the correlation increases and the error-squared decreases
until we reach $N\approx 4$ hidden nodes, after which both measures
level off. For the network with $N=8$ hidden nodes, a total of $87.8$
per cent of training data points and $85.4$ per cent of validation
points were correctly classified. A summary of the classification
results for this network is given Table~\ref{tab:cdatatrain}.
\begin{table}
\caption{Classification results for the converged NN with architecture
  $4+8+3$ for the Neal data set.}\label{tab:cdatatrain}
\begin{center}
\begin{tabular}{ccc|ccc}
\hline & \multirow{2}{*}{True class} & \multirow{2}{*}{Number} & \multicolumn{3}{|c}{Predicted class ($\%$)} \\
 & & & 0 & 1 & 2 \\ \hline
Training data & 0 & 282 & 84.0 & 4.96 & 11.0 \\
& 1 & 93 & 14.0 & 82.8 & 3.2 \\
& 2 & 386 & 7.0 & 1.3 & 91.7 \\ [2mm]
Validation data & 0 & 99 & 75.7 & 6.1 & 18.2 \\
& 1 & 19 & 21.1 & 78.9 & 0.0 \\
& 2 & 121 & 5.0 & 0.8 & 94.2 \\
\hline
\end{tabular}
\end{center}
\end{table}
These results compare well with Neal's own original results and are
similar to classifications based on applying the original criteria
directly to data points that have noise added.
Figure~\ref{fig:cdataplot} shows the data set and the
true classifications.

\subsection{Dimensionality reduction using autoencoders}
\label{sec:NNtoy_AE}

Dimensionality reduction is a very common task in astronomy, which is
usually performed using principal component analysis (PCA,
\citealt{kendall}) and its variants. In PCA, the eigenvalues and
eigenvectors of the correlation matrix of the centred data (from which
the mean has been subtracted) are found. The eigenvector directions
define a new set of variables that are mutually-orthogonal linear
combinations of the original variables describing each data
item. Dimensionality reduction is then achieved by keeping only those
combinations corresponding to (a certain number of) the largest
eigenvalues. PCA is limited, however, by its use of orthogonal
projections and this has led to more recent interest in independent
component analysis (ICA), which still constructs linear combinations
of the original variables, but relaxes the condition of orthogonality
(see, e.g., \citealt{Hyvarinen2000}). More specifically, ICA finds a
set of directions such that the projections of the data onto these
directions have maximum statistical independence, either by
minimisation of mutual information or maximization of non-Gaussianity.

As discussed in Section~\ref{sec:autoenc}, antoencoders provide a natural
generalisation of PCA and ICA, and constitute an intuitive approach to
non-linear dimensionality reduction that reduces to PCA in the special
case of a single hidden layer and linear activation functions.

\subsubsection{Multivariate Gaussian data}\label{sec:NNtoy_gauss}

To provide a quick comparison of autoencoders and traditional PCA, we
first consider two examples in which the data points are drawn from a
single multivariate Gaussian distribution, as assumed in PCA, using a
theoretical covariance matrix with given eigenvalues and eigenvectors.
As a basic check, in both cases we perform the main PCA step of
calculating the eigenvalues and eigenvectors of the sample covariance
matrix of the resulting data points, and we find that they match 
those assumed very closely.

In our first example, we draw the data points $(x_1,x_2)$ from a
two-dimensional correlated Gaussian. For this simple case, we first
train an autoencoder with a single hidden layer. Moreover, to effect a
dimensionality reduction, we place just one node in the hidden layer,
so the full network architecture is $2+1+2$. Whitening of the input
and output data using~\eqref{eq:whitening2} was performed. Pre-training was
also used as, even in
such a very small autoencoder network (with a total of 7 network
parameters), it is easy for the optimiser to fall into the large local
maximum where each output is the average of the inputs.

\begin{figure}
\begin{center}
\subfigure[]{\includegraphics[width=0.9\columnwidth]{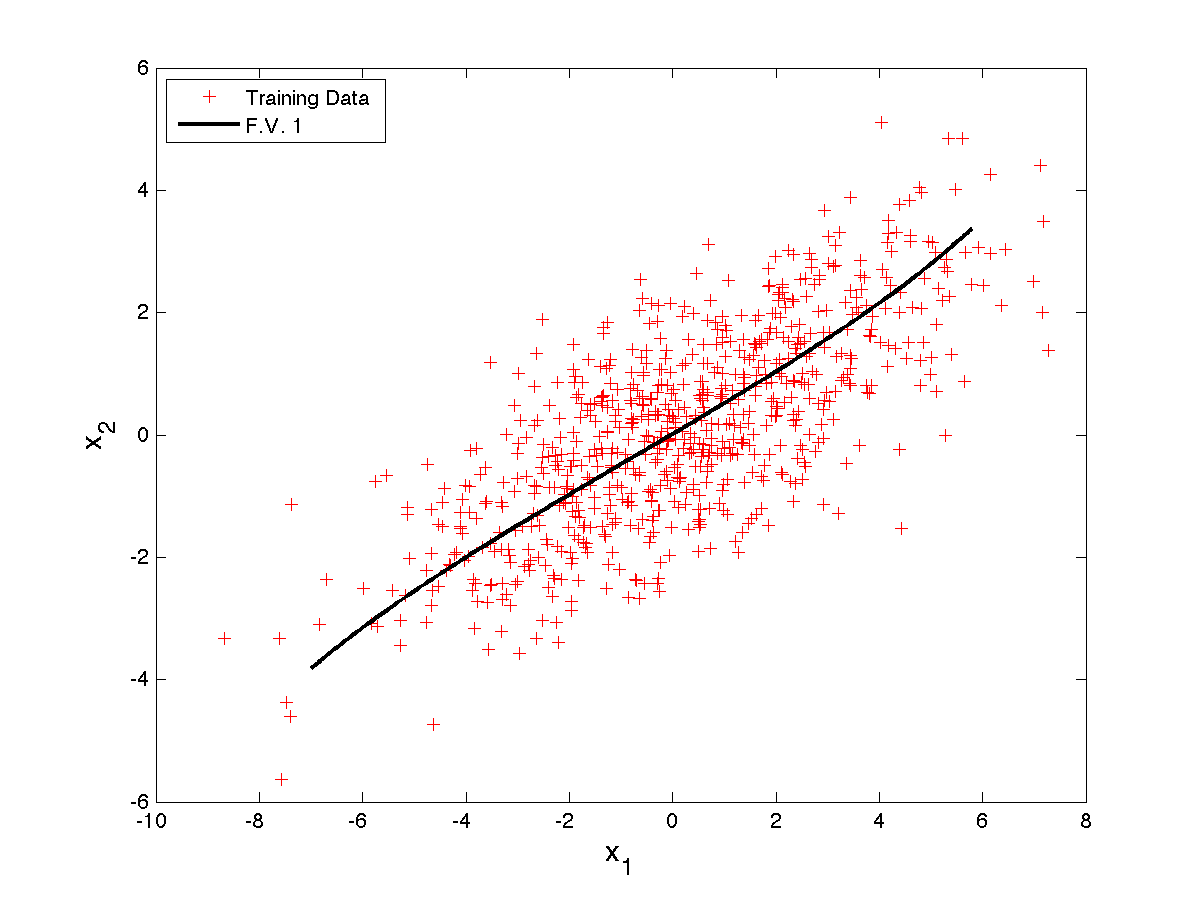}}
\subfigure[]{\includegraphics[width=0.9\columnwidth]{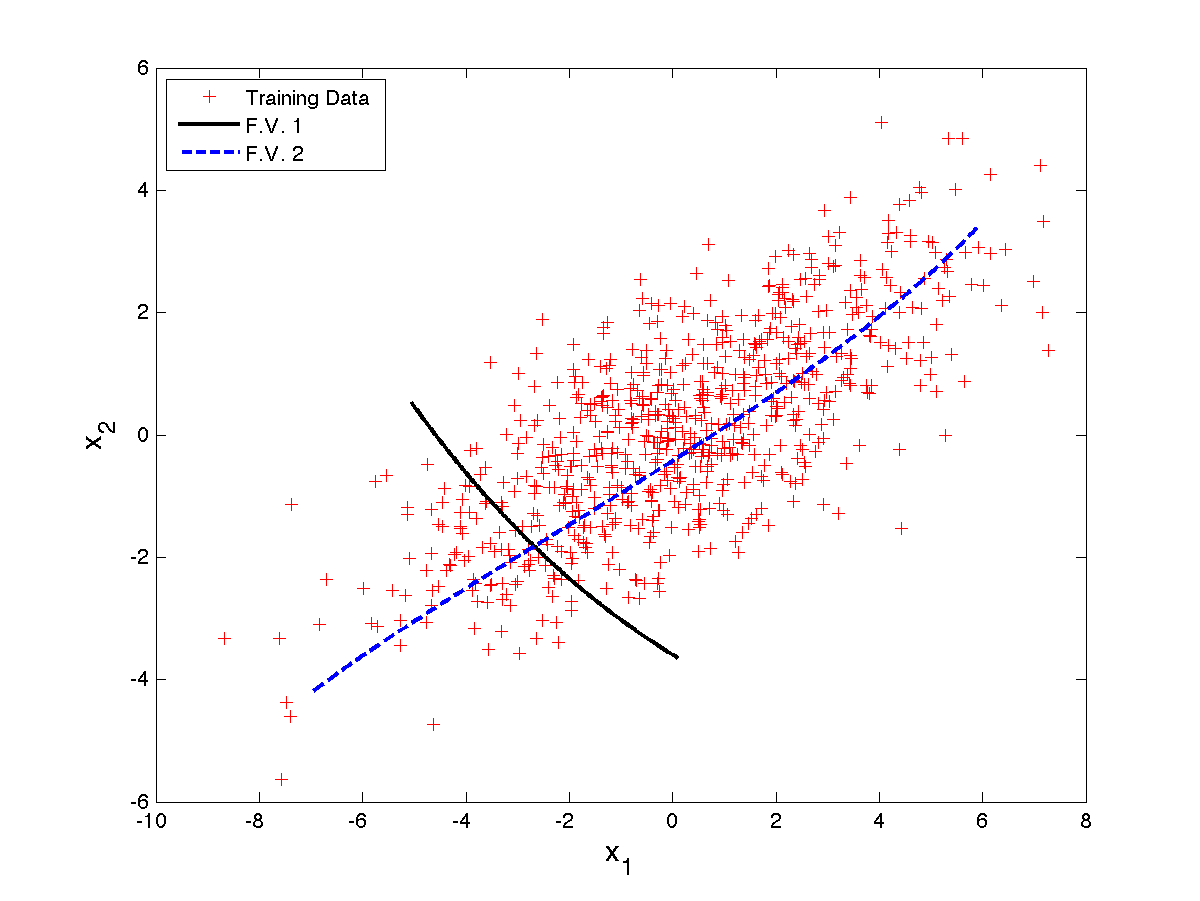}}
\caption{Original data points (red) drawn from a two-dimensional
  correlated Gaussian distribution, together with (a) the curve traced
  out by performing a decoding as one varies the single feature value
  $z_1$ in the central layer of a trained autoencoder with
  architecture $2+1+2$; and (b) two curves traced by performing a
  decoding as one varies the feature vectors $(z_1,0)$ and $(0,z_2)$,
  respectively, in the central layer of a trained autoencoder with
  architecture $2+2+2$. In both cases, the feature values are varied
  within the limits obtained when encoding the data.}
\label{fig:MVG-2Dfvs}
\end{center}
\end{figure}

Fig~\ref{fig:MVG-2Dfvs}(a) shows the original data and the curve
traced out in the data space when one performs a decoding as the value
of the feature variable $z_1$ in the single central layer node is
varied between the limits obtained when encoding the data. As one
might expect, this curve approximates the eigenvector with larger
eigenvalue of the covariance matrix of the data.  The curve is not
exactly a straight line because of the non-linearity of the activation
function from the hidden layer to the output layer, and is influenced
by the particular realisation of the data analysed. It should be noted
that dimensionality reduction is performed conversely, by
(non-linearly) encoding each data point $(x_1,x_2)$ to obtain the
corresponding feature value $z_1$ in the central layer node, rather
than performing any PCA-like (linear) projection in the data space.
The resulting error-squared and correlation for the antoencoder are
$0.476$ and $90.5$ percent, respectively.

To pursue the comparison with PCA further, we also train in a similar
manner an autoencoder with two nodes in a single hidden layer, so the
full network architecture is $2+2+2$, although this is clearly no
longer relevant for dimensionality reduction.
Fig~\ref{fig:MVG-2Dfvs}(b) again shows the original data, together
with the two curves traced out in the data space when one performs a
decoding as one varies (between the limits obtained when encoding the
data) the feature vectors $(z_1,0)$ and $(0,z_2)$, respectively, in
the central layer nodes. We see that, in the first case, one recovers
a curve very similar to that shown in Fig~\ref{fig:MVG-2Dfvs}(a),
whereas, in the second case, the curve approximates the eigenvector of
the data covariance matrix having the smaller eigenvalue. As before,
neither curve is exactly a straight line because of non-linearity of
the activation function. Moreover, the curves do not intersect at
right-angles, as would be the case for principal component
directions. The resulting correlation and error-squared for the
antoencoder are $0.022$ and $99.8$ percent, respectively. We note that
the latter is very close to $100$ percent, as one would expect for
this two-dimensional data set.

In our second example, we demonstrate the ability to determine the
optimal number of nodes in the single hidden layer (and hence the
optimal number of feature values) for an autoencoder, when redundant
information is provided. To accomplish this, we draw data points from
a 3-dimensional correlated Gaussian, but then `rotate' these points
into 5-dimensional space. Thus, each data point has the form
$(x_1,x_2,\ldots,x_5)$, but all the points lie on a 3-dimensional
hyperplane in this space. In a similar manner to above, we train
autoencoders with architecture $5+N+5$. As $N$ is varied, the
error-squared and correlation of the resulting networks are given in
Table~\ref{tab:MVG-5D}. As expected, once three hidden nodes are used,
the correlation is very close to $100$ percent, and adding more does
not improve the results.
\begin{table}
\caption{The error-squared and correlation for autoencoder networks
  with architecture $5+N+5$ applied to data points drawn from a
  3-dimensional correlated Gaussian distribution that are then
  `rotated' into a 5-dimensional space.}\label{tab:MVG-5D}
\begin{center}
\begin{tabular}{c|ccc}
\hline
$N_{\textrm{hid}}$ & Error-squared & Correlation $\%$ \\ \hline
1 & $0.00613$ & $79.6$ \\
2 & $0.00127$ & $96.0$ \\
3 & $4.87 \times 10^{-5}$ & $99.86$ \\
4 & $4.87 \times 10^{-5}$ & $99.86$ \\
5 & $4.87 \times 10^{-5}$ & $99.86$ \\
\hline
\end{tabular}
\end{center}
\end{table}

We note that, if desired, one can `rank' the feature variables
according to the amount by which they decrease the error-squared or
increase the correlation, which is analogous to ranking eigenvectors
according to their eigenvalues in PCA.

\subsubsection{Data distributed on a ring}\label{sec:NNtoy_autoring}

We now consider another two-dimensional example, but one for which the
data are drawn from a distribution very different to the single
multivariate Gaussian assumed by PCA. In particular, the data
$(x_1,x_2)$ are distributed about a partial ring centred on
$(0.5,0.5)$ with radius $0.5$ to produce a long curving degeneracy in
the data space. The data are generated according to
\begin{subequations}
\label{eq:ringdefn}
\begin{align}
\label{eq:rindefn_x}
x_1 &= 0.5 + (0.5-n)\cos\theta, \\
\label{eq:rindefn_y}
x_2 &= 0.5 + (0.5-n)\sin\theta,
\end{align}
\end{subequations}
where $\theta$ is drawn uniformly in the range $[0.1\pi,1.9\pi]$ and
$n$ is drawn from a Gaussian distribution with zero mean and standard
deviation of $0.1$; this example was originally presented
in~\cite{AstroNN}.

\begin{figure}
\begin{center}
\includegraphics[width=0.99\columnwidth]{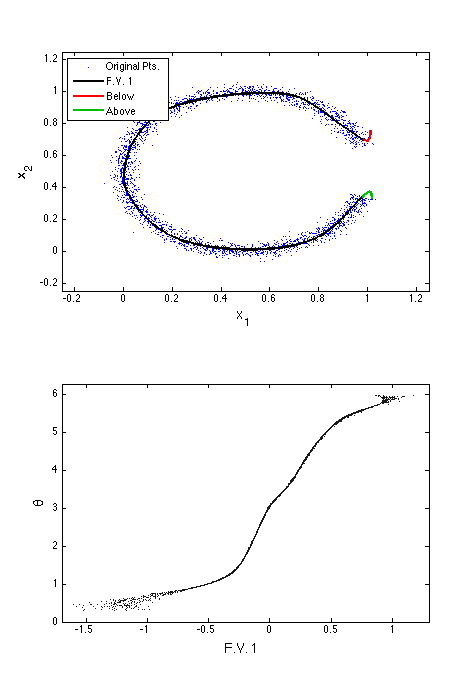}
\caption{(Top) Original data points (light blue) and the curve (black) traced
  out by performing a decoding as one varies (between the limits
  obtained when encoding the data) the single feature value $z_1$ in the
  central layer of a trained autoencoder with architecture
  $2+13+1+13+2$. The curve traced out when $z_1$ is allowed to vary below
  (above) the range encountered in training is shown in red (green).
  (Bottom) The true angle $\theta$ of the training data points versus their
  encoded feature values.}
\label{fig:ring}
\end{center}
\end{figure}

The data are plotted as the light blue points in Figure~\ref{fig:ring} (top).
Although the noiseless data are fully determined by the single
parameter $\theta$, it is clear that a linear dimensionality reduction
method, such as PCA, would be unable to represent this data set
accurately in a single component. Indeed, the dominant principal
component would lie along a straight, horizontal (symmetry) line
passing through the point $(0.5,0.5)$, as is easily verified in
practice. Projections onto this direction clearly do not
distinguish between data points having the same $x_1$-coordinate, but
lying on opposite sides of this symmetry line.  As we will now
show, however, it is possible to represent this data set well using
just a single variable, by taking advantage of the non-linearity of an
antoencoder.

\begin{figure}
\begin{center}
\includegraphics[width=0.95\columnwidth]{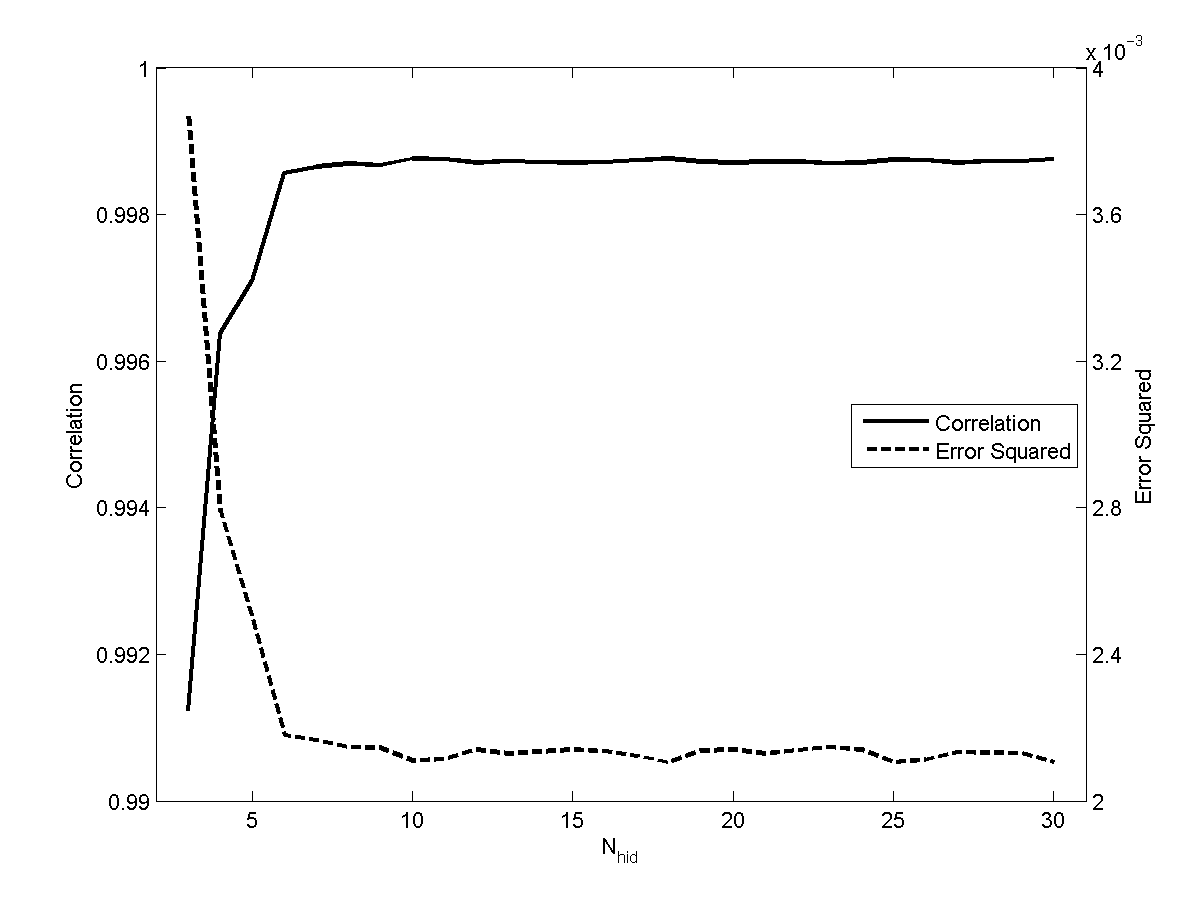}
\caption{The correlation and error-squared values as a function of the
  number of hidden nodes $N$ obtained from converged autoencoder with
  architecture $2+N+1+N+2$ for the data on a ring problem.}
\label{fig:ring_corr}
\end{center}
\end{figure}

In this slightly more challenging example, we train an autoencoder
with three hidden layers, again with a single node in the central
layer to perform a dimensionality reduction. Thus, the full network
architecture is $2+N+1+N+2$. Whitening of the input and output data
was applied using~\eqref{eq:whitening2}, and the network was pre-trained.  The
optimal value for $N$ is determined by comparing the correlation and
error-squared for networks with different numbers of hidden nodes.
These results are shown in Fig.~\ref{fig:ring_corr}, which shows
that optimal performance is reached for $N\ga 10$, beyond which no
significant improvement results from adding further hidden nodes.

The results obtained for the antoencoder with architecture
$2+13+1+13+2$ are shown in Figure~\ref{fig:ring}. The top panel shows
the curve (in black) traced out by performing a decoding as one varies
(between the limits obtained when encoding the data) the single
feature value $z_1$ in the central layer of a trained autoencoder with
architecture $2+13+1+13+2$; this clearly follows the ring structure
very closely. The curve traced out when $z_1$ is allowed to vary below
(above) the range encountered in training is shown in red (green).
One sees that each of these curves extend a short distance into the
gap in the ring, with one curve extending at either end of the
gap. Conversely, the bottom panel shows the encoded feature value
obtained as compared to the true angle $\theta$ of the input data. We
see that, as expected, there is a strong and monotonic relationship
between these two variables.

\subsubsection{Data distributed in multiple Gaussian modes}
\label{sec:NNtoy_automodes}

In this example, we again consider a two-dimensional case in which the
data are drawn from a distribution very different to a single
multivariate Gaussian, but, rather than simulating a long, curving
degeneracy, we focus here on a distribution possessing multiple modes.
In particular, the data are generated from the sum of four equal
Gaussian modes, each having a standard deviation of $0.1$ in both the
$x_1$ and $x_2$ directions, with means of $(0.25,0.25)$,
$(0.25,0.75)$, $(0.75,0.25)$, and $(0.75,0.75)$, respectively; this
example was also originally presented in~\cite{AstroNN}.

\begin{figure}
\begin{center}
\includegraphics[width=0.99\columnwidth]{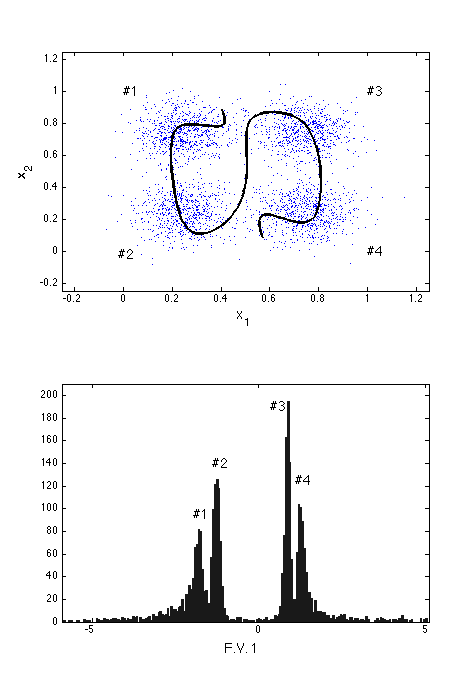}
\caption{(Top) Original data points (light blue) and the curve (black)
  traced out by performing a decoding as one varies (between the
  limits obtained when encoding the data) the single feature value
  $z_1$ in the central layer of a trained autoencoder with
  architecture $2+30+1+30+2$.  (Bottom) Histogram of the encoded feature
  values obtained from the input data; four separate peaks are
  visible, corresponding to the four Gaussian modes as labelled.}
\label{fig:modes}
\end{center}
\end{figure}

The data are plotted as the light blue points in
Figure~\ref{fig:modes} (top).  In this case it is not intuitively
obvious to what extent the data can be represented using a single
variable. It is once again clear, however, that a linear
dimensionality reduction method, such as PCA, would be unable to
represent this data set accurately in a single component. Indeed, in
this case, the two principal directions lie along diagonal (symmetry)
lines at $\pm 45$ degrees, passing through the point $(0.5,0.5)$, and
(in theory) have equal eigenvalues, so either direction can be used to
perform the dimensionality reduction.  Projection onto the line at
$+45$ degrees (say) will clearly not distinguish between any two data
points that lie on any given line perpendicular to that direction,
thereby conflating data points in modes 1 and 4 in the figure. Thus,
the resulting histogram of projected values for the data points will
contain only three (broad) peaks (see \citealt{AstroNN}). As we now
show, however, it is possible to distinguish all four modes using just
a single variable, if one again makes use of the non-linear nature of
autoencoders.

We again train an autoencoder with architecture $2+N+1+N+2$, using
whitening of the input with~\eqref{eq:whitening2}, and pre-training.
The optimal value for $N$ is determined by comparing the correlation
and error-squared for networks with different numbers of hidden
nodes. Investigating values for $N$ between $3$ and $30$, we find that
$N=30$ performed best by a small margin. The results from this
network are shown in Figure~\ref{fig:modes}. The top panel shows the
curve (in black) traced out by performing a decoding as one varies
(between the limits obtained when encoding the data) the single
feature value $z_1$ in the central layer of a trained autoencoder;
this curve traces out a path through the centre and outskirts of all
four modes, each of which corresponds to a distinct range of feature
values.  This is illustrated in the bottom panel, which shows the
histogram of the encoded feature values obtained from the data set.
The histogram contains four clear peaks, each corresponding to one of
the modes in the original data distribution.  By setting appropriate
threshold values on this feature variable, we can classify the points
as belonging to each of the modes with a $97.83$ per cent accuracy
(classifying based on the raw $x_1$ and $x_2$ values yields only a
slightly better accuracy of $98.85$ per cent). 

\subsection{MNIST handwriting recognition}\label{sec:MNIST}

\begin{figure}
\begin{center}
\includegraphics[width=\columnwidth]{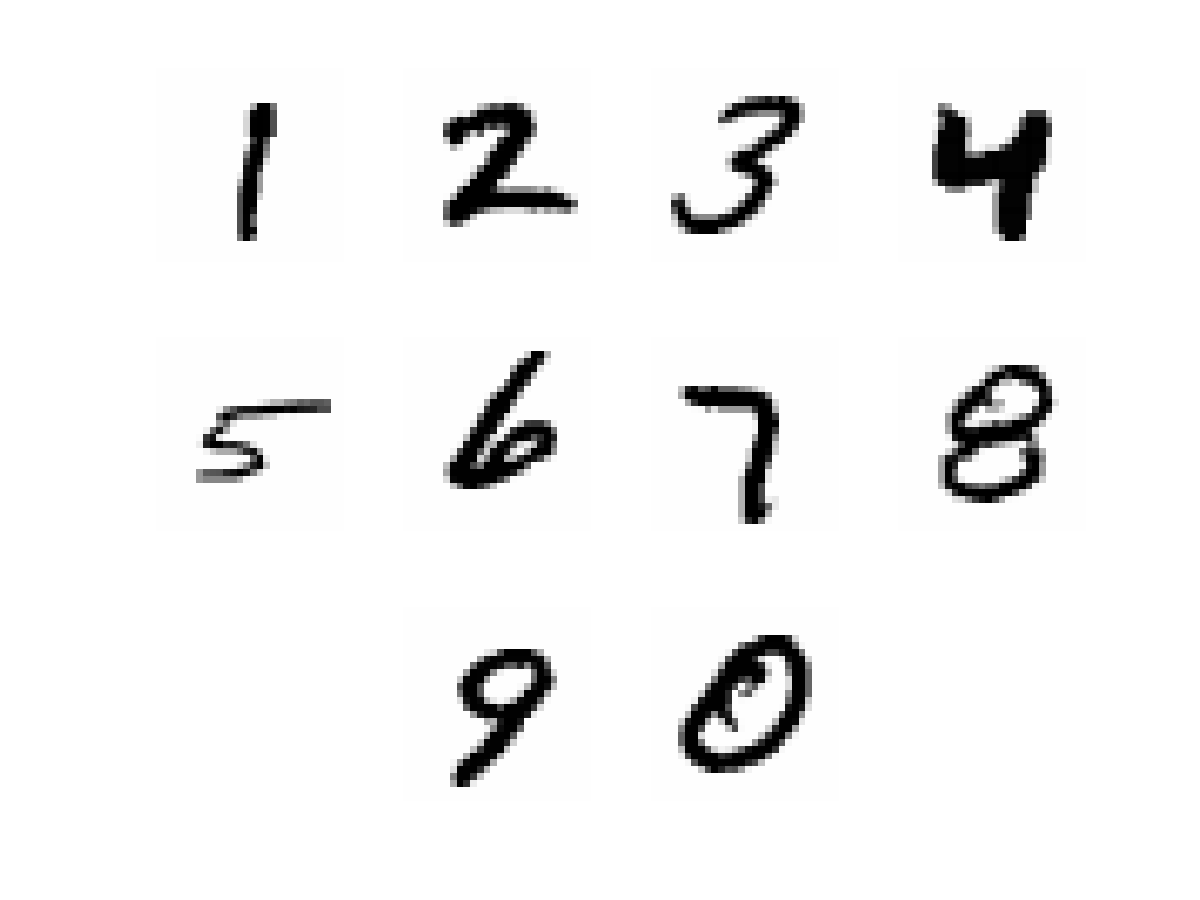}
\caption{Sample handwritten digits from the MNIST database.}
\label{fig:MNISTsamples}
\end{center}
\end{figure}

The MNIST database of handwritten digits is a subset of a larger
collection available from NIST (National Institute for Standards and
Technology). It consists of $60,000$ training and $10,000$ validation
images of handwritten digits. Each digit has been size-normalised and
centred in a $28 \times 28$ pixel greyscale image. The images are
publicly available
online~\footnote{\texttt{http://yann.lecun.com/exdb/mnist/}}, along
with more information on the generation of the data set and results
from previous analyses by other researchers. This data set has become
a standard for testing of machine learning algorithms.  Some sample
digits are shown in Figure~\ref{fig:MNISTsamples}.  The learning task
is to identify correctly the digit written in each image.  Although
this may be an easy task for a human brain, it is quite a challenging
machine learning application.

\subsubsection{Direct classification}

Several classification networks of varying complexity were trained on
this problem. In all cases, pre-training was used for any hidden
layers and all inputs were whitened using the transformation
\eqref{eq:whitening2}.  Each network has $28 \times 28 = 784$ inputs 
(corresponding to the greyscale image size) and $10$ outputs (one for
each digit). The class (digit) assigned to each image was that having
the highest output probability.

The results obtained are summarised in Table~\ref{tab:MNISTresults},
where the error rates, defined as the fraction incorrectly classified, are those
calculated on the set of validation images. We note that some of these
networks are large and deep, but are nonetheless well trained using
{\sc SkyNet}. These results may be compared with those referenced on
the MNIST website, which yield error rates as low as $0.35$ percent~\citep{MNISTdeepsoln} but more typically between $1$ and $5$ percent~\citep{MNISTlecun98}.
\begin{table}
\caption{Error rates for classification networks with different
  architectures, trained to identify handwritten digits in greyscale
  images from the MNIST database. All networks have $784$ inputs and
  $10$ outputs.}
\label{tab:MNISTresults}
\begin{center}
\begin{tabular}{c|c}
\hline
Hidden layers  & Error rate ($\%$) \\ \hline 
0 & 8.08 \\
100 & 4.15 \\
250 & 3.00 \\
1000 & 2.38 \\
$300+30$ & 2.83 \\
$500+50$ & 2.62 \\
$1000+300+30$ & 2.31 \\
$500+500+2000$ & 1.76 \\
\hline
\end{tabular}
\end{center}
\end{table}

\subsubsection{Dimensionality reduction and classification}

A dimensionality reduction of the MNIST data was also performed. In
particular, two large and deep autoencoder networks were trained, one
with hidden layers $1000+300+30+300+1000$ (called AE-30) and the other
with hidden layers $1000+500+50+500+1000$ (called AE-50); both
networks have $784$ inputs and outputs, corresponding to the image
size. Thus, the dimensionality reduction to $30$ or $50$ feature
variables, respectively, represents a significant data compression.
The AE-30 network was able to obtain an average total error squared of
only $4.64$ and AE-50 obtained an average total error squared of
$3.29$. These values are comparable to those obtained
in~\cite{Hinton_AE_Science}. Thus, despite reducing the dimensionality
of the input data used from $784$ pixels to $30$ or $50$ non-linear
feature variables, these reduced basis sets retain enough information
about the original images to reproduce them to within small
errors. This also demonstrates that {\sc SkyNet} is capable of
training large and deep autoencoder networks.

As mentioned previously, dimensionality reduction is sometimes used as
a prelude to a supervised-learning task such as classification, since
the latter can sometimes be performed just as accurately (or sometimes
more so) in the reduced space as in the original data space. To
illustrate this approach, using each of AE-30 and AE-50, all the
training images were passed through the autoencoder to obtain the
($30$ or $50$) encoded feature values for each image. These feature
values (rather than the original images) were then used to train a
classification network (with just $30$ or $50$ input nodes,
respectively, and $10$ output nodes) to identify the handwritten
digits. The resulting error rates are listed in
Table~\ref{tab:MNISTresultsAE}, for networks with different numbers of
hidden layers and nodes.
\begin{table}
\caption{Error rates for classification networks with different
  architectures, trained on autoencoder feature values to identify
  handwritten digits from the MNIST database. For AE-30 (AE-50), all the
  classification networks have 30 (50) inputs and 10 outputs.}
\label{tab:MNISTresultsAE}
\begin{center}
\begin{tabular}{c|c|c}
\hline
Autoencoder & Classification hidden layers & Error rate ($\%$) \\ \hline
\multirow{4}{*}{AE-30} & $0$ & 9.57 \\
 & $10$ & 6.39 \\
 & $30$ & 3.03 \\
 & $100+50+10$ & 2.55 \\ \hline
\multirow{4}{*}{AE-50} & $0$ & 8.68 \\
 & $10$ & 6.61 \\
 & $50$ & 2.65 \\
 & $100+50+10$ & 2.71 \\
\hline
\end{tabular}
\end{center}
\end{table}
In particular, comparing with Table~\ref{tab:MNISTresults}, we see
that the resulting classifications are nearly as accurate as the
best-performing network trained on the full images, despite reducing
the dimensionality of the input data from $784$ pixels to $30$ or $50$
non-linear feature variables, and reducing the number of
classification network parameters by several orders of magnitude.

\subsubsection{Dimensionality reduction and clustering}

Finally, a massive dimensionality reduction to just two feature
variables was performed on the images by training a very large and
deep autoencoder, with hidden layers $1000+500+250+2+250+500+1000$
(and $784$ inputs and outputs). As expected, this network is
significantly less able to reproduce the original images, having an
average total error squared of $31.0$, but has the advantage that one
can plot the two feature values obtained for each of the images to
provide a simple illustration of clustering.

\begin{figure}
\begin{center}
\includegraphics[width=\columnwidth]{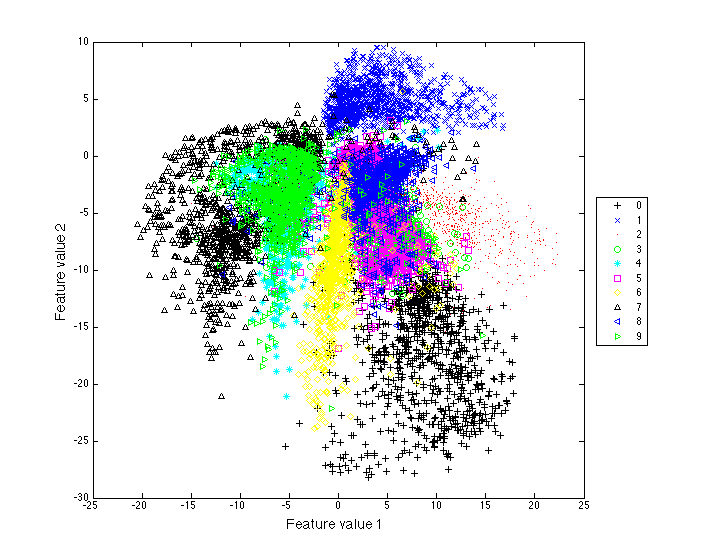}
\caption{Scatterplot of the two feature variables for the $10,000$
  validation images from the MNIST database, obtained using the
  encoding half $784+1000+500+250+2$ of a symmetric autoencoder.}
\label{fig:MNIST_fvclasses}
\end{center}
\end{figure}

Such a scatterplot is shown in Fig.~\ref{fig:MNIST_fvclasses} for the
$10,000$ validation images, where the points are colour-coded
according to the true digit contained in each image; this figure may
be compared with figure 3B in~\cite{Hinton_AE_Science}. We see that
there is some significant overlap between digits with similar shapes,
but that some digits do occupy distinct regions of the parameter space
(particularly $1$ in the top right, some $0$s in the bottom right, and
many examples of $2$ in the middle right).

\subsection{Comparison with FANN library}

In this section, we perform a simple comparison between {\sc SkyNet} and an
alternative algorithm for training a NN. The case we use is the simple sinc problem from
Section~\ref{sec:NNtoy_sinc} and we compare against the FANN library. This is a NN library
that has been developed over several years and thus has more features and interfaces than implemeneted
thus far for {\sc SkyNet}. However, the training is performed via standard backpropagation techniques
which are first-order in nature.

Training on the exact same sinc data with one hidden layer containing $7$ nodes,
we found that FANN's optimal predictions and run time were equivalent
to those of {\sc SkyNet}. However, FANN performed approximately an order of magnitude more steps in parameter
space to reach this solution (meaning an individual step was similarly faster) and had no feature to prevent
overfitting had a larger NN been trained (uses only target error and maximum number of steps).
Furthermore, while running multiple times with {\sc SkyNet}
produces similar run time and results consistently, FANN run times varied greatly and often did not converge to the same
result. Lastly, FANN requires the user to create their own ``main'' function in C (or another language)
to setup the network to be trained, read in data, perform training, and save the network. By comparison,
{\sc SkyNet} seeks to make these functions easier for the user by asking only for an input settings file
and formatted data. The additional functionality can be implemented in future releases while a useful and simple
tool is provided now.

\section{Applications:  Astrophysical Examples}\label{sec:NNex_astro}

\subsection{Regression: Mapping Dark Matter challenge}\label{sec:MDM}

\begin{figure}
\begin{center}
\subfigure[]{\includegraphics[width=0.4\columnwidth]{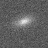}}
\subfigure[]{\includegraphics[width=0.4\columnwidth]{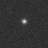}}
\caption{Example image pair of (a) galaxy and (b) star from the
  Mapping Dark Matter Challenge; each image contains $48\times48$
  greyscale pixels.}
\label{fig:MDMexample}
\end{center}
\end{figure}

The Mapping Dark Matter (MDM) Challenge was presented on
\texttt{www.kaggle.com} as a simplified version of the GREAT10
Challenge~\citep{GREAT10}. In this problem, each data item consists of
two $48 \times 48$ greyscale images of a galaxy and a star,
respectively. Each pixel value in each image is drawn from a Poisson
distribution with mean equal to the (unknown) underlying intensity
value in that pixel. Moreover, both images have been convolved with
the same, but unknown, point spread function.  The learning task is,
for each pair of images, to predict the ellipticities $e_1$ and $e_2$
(defined below) of the underlying true galaxy image, and thus
constitutes a regression problem.  The training data set contains
$40,000$ image pairs and the challenge data set contains
$60,000$ image pairs. A sample galaxy and star image pair is shown in
Fig~\ref{fig:MDMexample}.
During the challenge, the solutions for the validation data set were
kept secret and participating teams submitted their
predictions. Further details about the challenge and descriptions of
the top results can be found in~\cite{MDMresults}.

\subsubsection{Galaxy image model}\label{sec:MDMdata}

\begin{figure}
\begin{center}
\includegraphics[width=0.8\columnwidth]{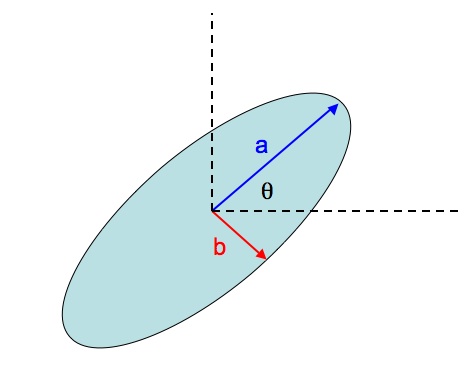}
\caption{Definition of the underlying galaxy image ellipse parameters
  used in the Mapping Dark Matter Challenge.  Image from
  \texttt{http://www.kaggle.com/c/mdm/}.}
\label{fig:ellipticity}
\end{center}
\end{figure}

The true underlying galaxy image is assumed to be elliptical with
semi-major axis $a$, semi-minor axis $b$, and position
angle $\theta$, as shown in Fig.~\ref{fig:ellipticity}.
The ellipticities $e_1$ and $e_2$ are related to these parameters by
\begin{subequations}
\label{eq:ellipticity}
\begin{align}
\label{eq:ellipticity1}
e_1 &= \frac{a-b}{a+b} \cos(2 \theta), \\
\label{eq:ellipticity2}
e_2 &= \frac{a-b}{a+b} \sin(2 \theta),
\end{align}
\end{subequations}
and may vary in the range $[-1,1]$. Further details about the data set
can be found at the challenge's
webpage~\footnote{\texttt{http://www.kaggle.com/c/mdm/}}. This also
gives the unweighted quadrupole moments (UWQM) formula for calculating
the ellipticities of an image. As the competition organisers note,
however, this formula will not provide very good esimates of the true
galaxy ellipticities, since it does not account for the point-spread
function or noise.

\subsubsection{Results}\label{sec:MDMresults}

We use {\sc SkyNet} to train several regression networks, each of
which takes the galaxy and star images as inputs and produces
estimates of the true galaxy ellipticities as outputs.  Following the
approach used in the original challenge, the quality of a network's
predictions are measured by the root mean squared error (RMSE) of its
predicted ellipticities over the challenge data set of $60,000$ pairs
of images. Clearly, better predictions result in lower values of the
RMSE.

The size of the dataset meant that training large networks was
very computationally expensive. Therefore, for this demonstration, we
train only relatively small networks, but used three different data
sets: (i) the full galaxy and star images; (ii) the full galaxy image
and a centrally cropped star image; and (iii) the full galaxy image
alone. Of the training data provided, consisting of $40,000$ image
pairs, $75$ per cent were used for training the networks (without
pre-training and whitening using transformation \eqref{eq:whitening2})
and the remaining $25$ percent were used for validation. 
The RMSE values for trained networks of
different architecture, evaluated on the challenge
dataset, are given in Table~\ref{tab:MDMresults}.
\begin{table}
\caption{Root mean square errors on ellipticity predictions for
  networks with different architectures, evaluated on the $60,000$
  image pairs of the Mapping Dark Matter Challenge. All networks have
  two outputs: the galaxy ellipticities $e_1$ and $e_2$.}
\label{tab:MDMresults}
\begin{center}
\begin{tabular}{l|cc}
\hline
Data set & Hidden layers & RMSE \\ \hline 
Full galaxy and star images & 0 & 0.0224146 \\
($48 \times 48 \times 2 = 4608$ inputs) & 2 & 0.0186944 \\
& 5 & 0.0184237 \\
& 10 & 0.0182661 \\ [2mm]
Full galaxy and cropped star images & 0 & 0.0175578 \\
($48\times 48 + 24 \times 24 = 2880$ inputs)
& 2 & 0.0176236 \\
& 5 & 0.0175945 \\
& 10 & 0.0174997 \\
& 50 & 0.0172977 \\
& 50+10 & 0.0171719 \\ [2mm]
Full galaxy image only & 0 & 0.0234740 \\
($48 \times 48 = 2304$ inputs)
& 2 & 0.0234669 \\
& 5 & 0.0236508 \\
& 10 & 0.0226440 \\
\hline
\end{tabular}
\end{center}
\end{table}

The RMSE values obtained even for the naive first approach of using
the full dataset (i) as inputs are quite good; for comparison, the
standard software package {\sc SExtractor}~\citep{SExtractor} produced an
RMSE score of $0.0862191$ on this test data and the UWQM method scores
$0.1855600$. One see from Table~\ref{tab:MDMresults} that increasing
the number of hidden nodes beyond two does improve the network
accuracy, but only very slowly.

The NN results can, however, be improved more easily simply by
reducing the number of inputs without affecting the information
content, for example by cropping the star images to the central
$24\times24$ pixels to yield dataset (ii). This simple change
increases the accuracy of the ellipticity predictions, thereby
lowering the RMSE. Increasing the number of nodes in a single hidden
layer, or adding an extra hidden layer, does yield improving
predictions, although the rate of improvement is quite gradual.
Nonetheless, this indicates that more complex networks could further
improve the accuracy of the ellipticity predictions. The best result
obtained for the networks investigated, with an RMSE of $0.0171719$,
compares well with the competition winners, who achieved an RMSE of
$0.0150907$~\citep{MDMresults} using a mixture of methods that included
NNs. We note that our score, produced using an immediate
`out-of-the-box' application of {\sc SkyNet} that involves no
specialised data processing, would have placed us in $32$nd place out
of $66$ teams that entered the challenge.

We see from Table~\ref{tab:MDMresults}, however, that reducing the
number of inputs further by removing the star images altogether leads
to a significant increase in the RMSE. This is to be expected since
the absence of the star images does not allow for the NN to infer the
point-spread function sufficiently well to predict the underlying
galaxy ellipticities accurately.

Finally, we note that all of our results could potentially be improved
further by fitting profiles to the images and using the parameters of
these fits for training, which would reduce the number of inputs by
about two orders of magnitude. Alternatively, one could train an
autoencoder to dimensionally-reduce each image into a set of feature
variables; this would again vastly reduce the number of network inputs
also potentially alleviate the effect of noise in the images. Such
investigations are, however, postponed to a future publication. 

\subsection{Classification: identifying gamma-ray bursters}\label{sec:GRB}


Long gamma-ray bursts (GRBs) are almost all indicators of
core-collapse supernovae from the deaths of massive stars. The ability
to determine the intrinsic rate of these events as a function of
redshift is essential for studying numerous aspects of stellar
evolution and cosmology. The Swift space telescope is a
multi-wavelength detector that is currently observing hundreds of GRBs
\citep{Gehrels2004}.  However, Swift uses a complicated combination of
over 500 triggering criteria for identifying GRBs, which makes it
difficult to infer the intrinsic GRB rate. Indeed, most studies
approximate the Swift trigger algorithm by a single detection
threshold, which can lead to biasses in inferring the intrinsic GRB
rate as a function of redshift.

To investigate this issue further, a recent study by \cite{LienGRB}
performed a Monte Carlo analysis that generated a mock sample of GRBs,
using the GRB rate and luminosity function of \cite{Wanderman2010},
and processed them through an entire simulated Swift detection
pipeline, applying the full set of Swift trigger criteria, to
determine which GRBs would be detected. It was found that the
resulting measured GRB rate as a function of redshift followed very
closely that of the true Swift GRB set described in \cite{Fynbo2009};
this finding is consistent with both the mock GRB sample and the
simulated trigger pipeline being good approximations to reality.  This
analysis was, however, quite computationally expensive, since
determining if each GRB is detected by Swift requires over a minute on
a single CPU.

\subsubsection{Form of the classification problem}\label{sec:GRBproblem}

Our goal here is to replace the simulated Swift trigger pipeline with
a classification NN, which (as we will see) can determine in just a
few microseconds whether a given GRB is detected.  To this end, we use
as training data a pre-computed mock sample of $10,000$ GRBs from
\cite{LienGRB}. In particular, we divide this sample randomly into
$\sim 4000$ for training, $\sim 1000$ for validation, and the final
$\sim 5000$ as a blind test set on which to perform our final
evaluations. For each GRB we use use 13 inputs: the GRB total
luminosity, redshift, and energy peak, together with the arrival bin
at the Swift detector, bin size of the light curve in the emitting
frame, $\alpha$ and $\beta$ parameters for the GRB's Band function
spectrum, background intensity in four different energy ranges
($15$-$25$~keV, $15$-$50$~keV, $25$-$100$~keV, and $50$-$350$~keV),
angle of arrival at the detector, and total GRB flux. The two
softmaxed outputs correspond to probabilities (which sum to unity) for
the class $0$, that the GRB is not detected, and for the class $1$,
that the GRB is detected. In our analysis, we will focus on the
probability for class 1.

Different NN architectures were trained on these data and it was found
that NNs with hidden layer configurations of $50$, $100+30$, and
$300+100+30$ all performed equally well on the classification
task. Thus results presented in this section are those from the
network with hidden layers $100+30$ applied to the blind set of
$\sim 5000$ GRBs.

\subsubsection{Results}\label{sec:GRBresults}

Since the NN outputs are probabilities, we can investigate the quality
of the classification as a function of the threshold probability,
$p_{\textrm{th}}$, required in class 1 to claim a detection.  As
discussed in~\cite{Feroz2008}, we can compute the {\em expected}
number of total GRB detections, correct detections, and false
detections, as well as other derived statistics as a function of
$p_{\textrm{th}}$, {\em without} knowing the true classifications (as
would be the case in the analysis of real data).  If we label the
probability of detection for each GRB in the blind set as $p_i$, then
the expected total number $\hat{N}^{\textrm{total}}$ of GRBs, expected
number correctly predicted, $\hat{N}^{\textrm{true}}$, and expected
number falsely predicted, $\hat{N}^{\textrm{false}}$, are given by the
following:
\begin{subequations}
\label{eq:expectedNums}
\begin{align}
\label{eq:expectedTotal}
\hat{N}^{\textrm{total}} &= \sum_{i=1}^N p_i, \\
\label{eq:expectedTrue}
\hat{N}^{\textrm{true}} &= \sum_{i=1,p_i \geq p_{\textrm{th}}}^N p_i, \\
\label{eq:expectedFalse}
\hat{N}^{\textrm{false}} &= \sum_{i=1,p_i \geq p_{\textrm{th}}}^N 1-p_i,
\end{align}
\end{subequations}
where $N$ is the total number in the blind sample. From this, we can
compute the completeness $\epsilon$ (fraction of detected GRBs that
have been correctly classified; also referred to as the efficiency)
and purity $\tau$ (fraction of all GRBs that have been correctly
classified as detected), which are given by
\begin{subequations}
\label{eq:classdfns}
\begin{align}
\label{eq:completeness}
\epsilon &= \frac{\hat{N}^{\textrm{true}}}{\hat{N}^{\textrm{total}}} \\
\label{eq:purity}
\tau &= \frac{\hat{N}^{\textrm{true}}}{\hat{N}^{\textrm{true}} + \hat{N}^{\textrm{false}}}
\end{align}
\end{subequations}

\begin{figure}
\begin{center}
\includegraphics[width=0.9\columnwidth]{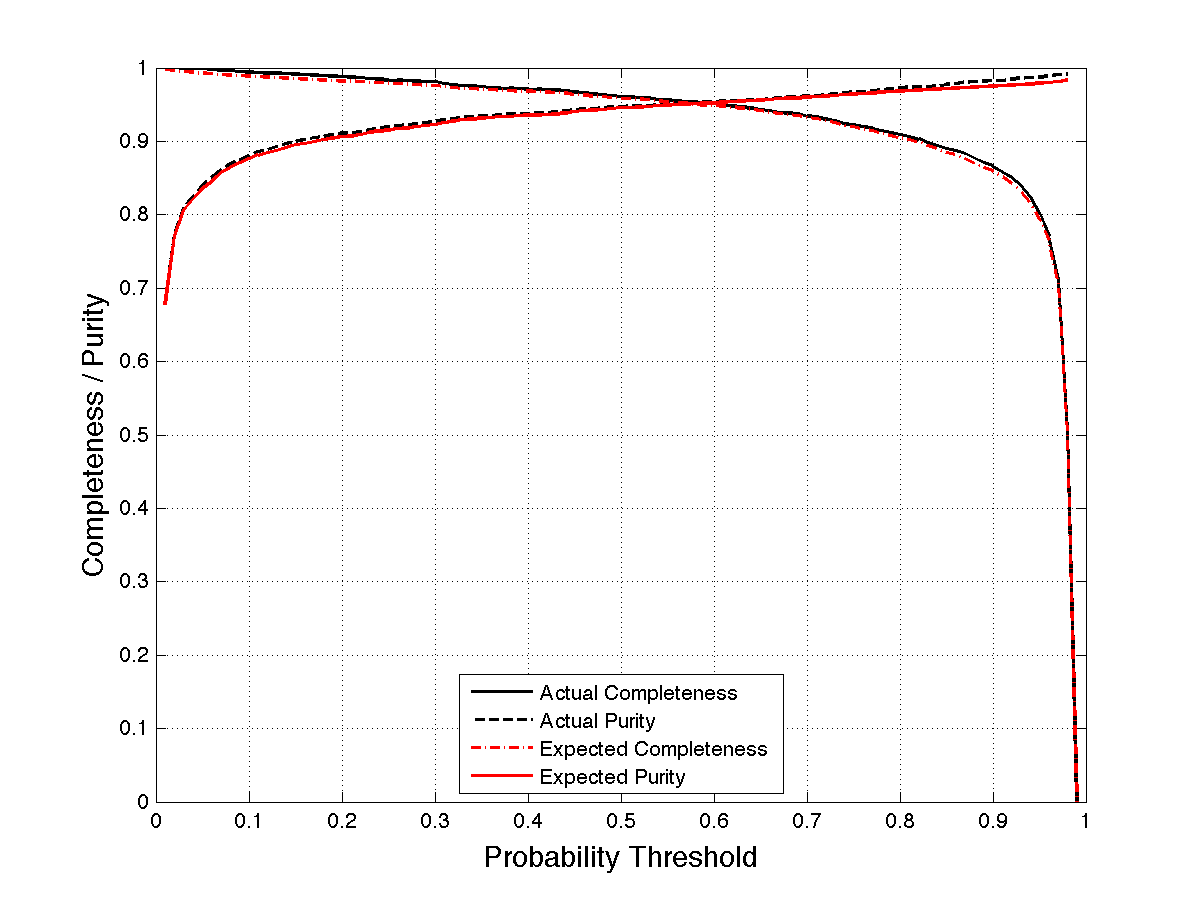}
\caption{Actual and expected values for the completeness and purity as
  a function of probability threshold, $p_{\textrm{th}}$. The expected
  curves are very accurate predictors of the actual ones.}
\label{fig:GRBcomppur}
\end{center}
\end{figure}

In Figure~\ref{fig:GRBcomppur} we plot the actual and expected
completeness and purity. It is clear that the actual and expected
curves lie on top of one another with only minimal differences. Thus,
without knowing the true classifications of the GRBs as detected or
not, we can set $p_{\textrm{th}}$ to obtain the desired completeness
and purity levels for the final sample.

With this information, we can also plot the actual and expected
receiver operating characteristic (ROC) curves (see, e.g.,
\citealt{Fawcett2006}). The ROC curve originated in signal detection
theory and is a reliable way of choosing an optimal threshold value as
well as comparing binary classifiers.  The ROC curve plots the true
positive rate (identical to completeness and also equal to the
Neyman--Pearson `power' of a test) against the false positive rate
(also known as contamination and the Neyman--Pearson type-I error
rate) as a function of the threshold value.\footnote{It is worth
  noting that, in terms of conditional probabilities, completeness is
  simply $\Pr(\mbox{classified as detected}|\mbox{detected})$, purity is its Bayes’
  theorem complement $\Pr(\mbox{detected}|\mbox{classified as detected})$, and
  contamination is $\Pr(\mbox{classified as detected}|\mbox{not detected})$.} A
perfect classifier will have a ROC curve that connects $(0,0)$ to
$(0,1)$ and then $(1,1)$, whereas a random classifier will yield a ROC
curve that is the diagonal line connecting $(0,0)$ and $(1,1)$
directly. In general, the larger the area underneath a ROC curve, the
more powerful the classifier. 

\begin{figure}
\begin{center}
\includegraphics[width=0.9\columnwidth]{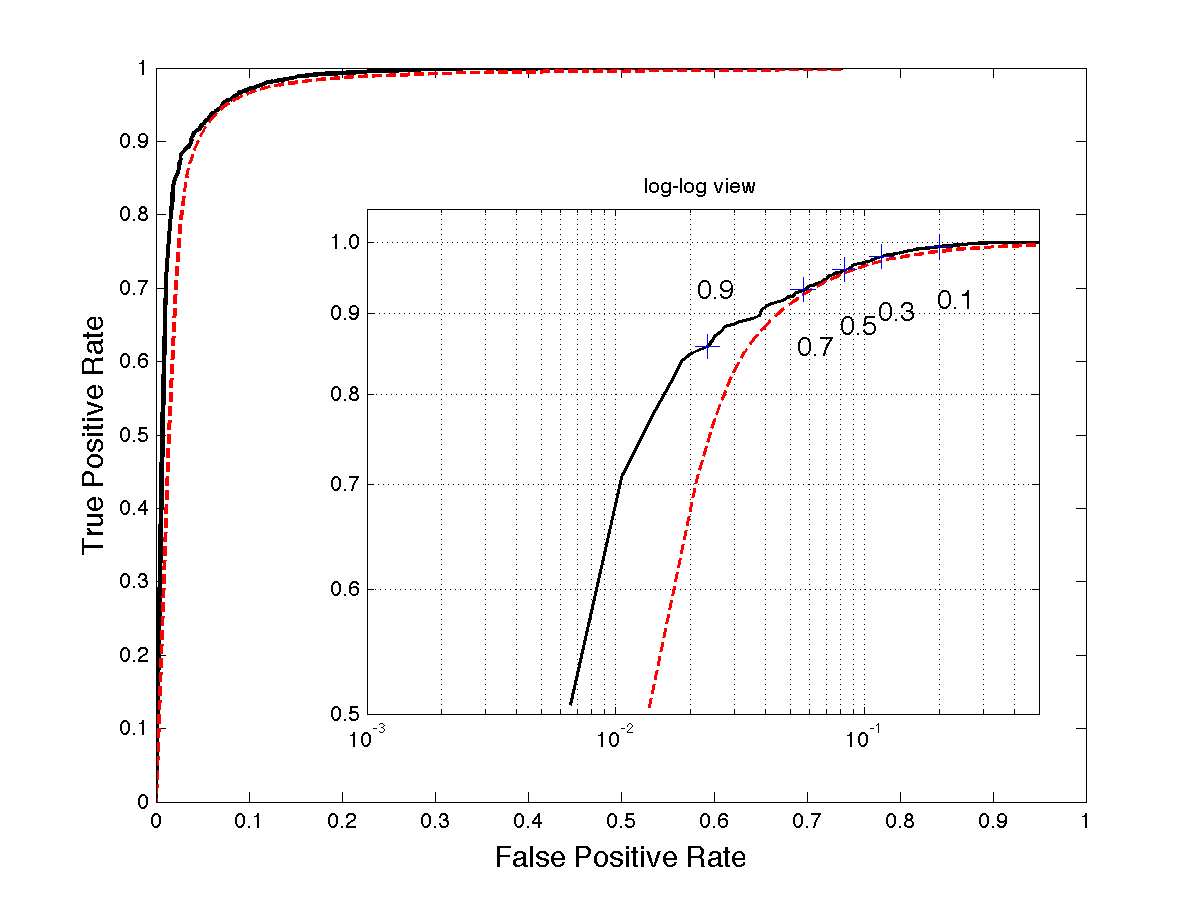}
\caption{Actual (black solid) and expected (dashed red) ROC curves for
  a NN classifier that predicts whether a GRB will be detected by
  Swift.  The curve traces true versus false positive rates as the
  probability threshold varies, as illustrated on the inset log-log
  plot.}
\label{fig:GRBroc}
\end{center}
\end{figure}

From Figure~\ref{fig:GRBroc}, we can see that the expected and actual
ROC curves for a NN classifier are very close, with small deviations
occuring only at very low false positive rates; it should be noted
that here the actual ROC curve is better than the expected one. The
curves also indicate that this test is quite powerful at
predicting which GRBs will be detected by Swift.

Using the completeness, purity, and ROC curves, one can make a
decision as to appropriate value of $p_{\textrm{th}}$ to use. One may
require a certain level of completeness, regardless of false
positives, or we may require a minimal level of contamination in the
final sample. Alternatively , one can use the ROC curve to derive an
optimal value for $p_{\textrm{th}}$.  For example, one can use the
$p_{\textrm{th}}$ value where the ROC curve intersects with the
diagonal line connecting $(0,1)$ and $(1,0)$, or where the line from
the point $(1,0)$ intersects the ROC curve at right angles. From
either of these criteria, we conclude that $p_{\textrm{th}}=0.5$, the
original naive choice, is a near-optimal threshold value.

\begin{figure}
\begin{center}
\includegraphics[width=0.85\columnwidth]{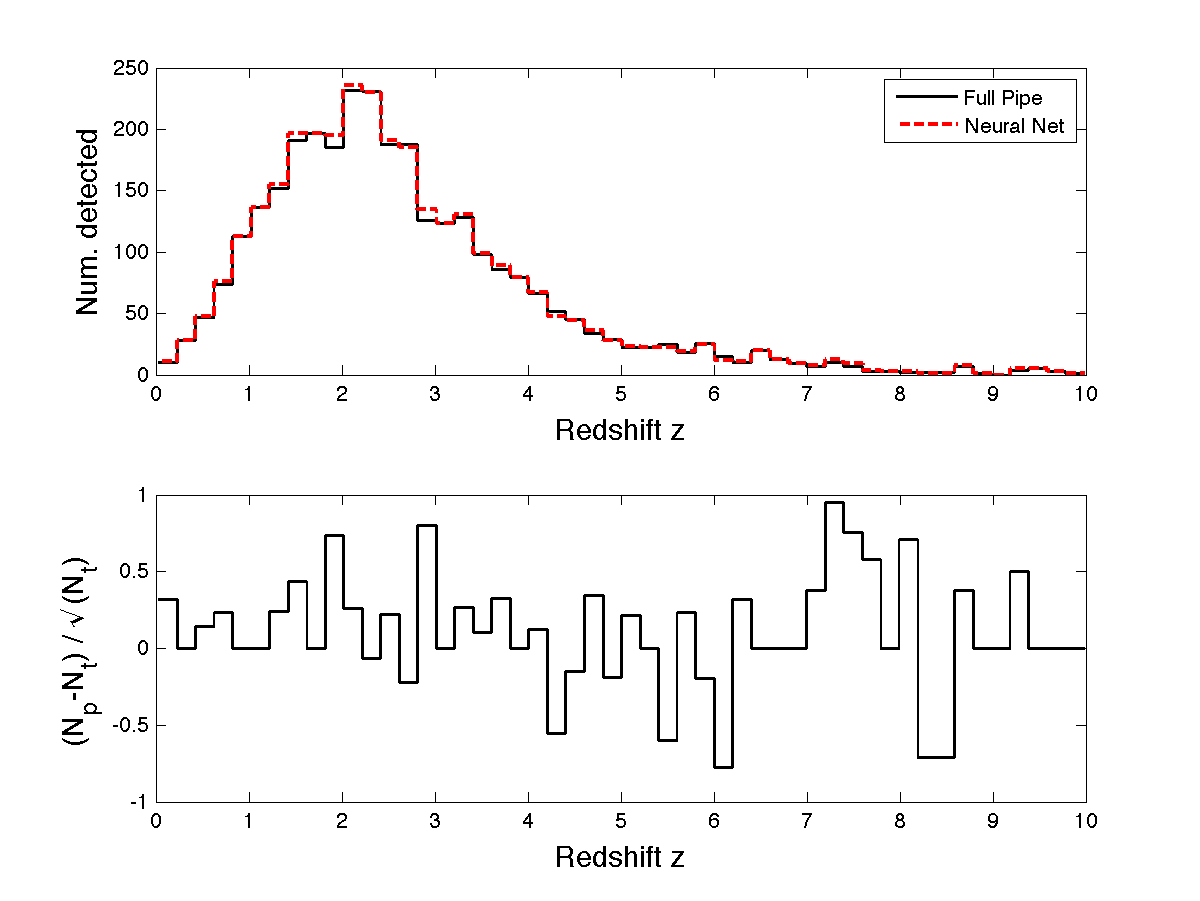}
\caption{(Top) Comparison of the detected GRB event counts obtained
  using the full simulated Swift trigger pipeline and our
  classification NN.
(Bottom) Normalized error between the two counts.}
\label{fig:GRBzdist}
\end{center}
\end{figure}

Using $p_{\textrm{th}}=0.5$, we now wish to determine how well the GRB
detection rate with respect to redshift is reproduced, since this
relationship is key to deriving scientific results. In
Figure~\ref{fig:GRBzdist}, we show the event counts as a function of
redshift for both our NN classifier and the original simulated Swift
trigger pipeline. It is clear that the two sets of counts agree very well.
In the bottom part of the figure, we
calculate a measure of the error within each redshift bin by computing
$(N_{\rm t}-N_{\rm p})/\sqrt{N_{\rm t}}$, where $N_{\rm t}$ is the `true' 
number of GRBs detected  by the full simulated Swift pipeline, and
$N_{\rm p}$ is the number obtained using our classification NN.
As the original detection counting is
essentially a Poissonian process, its intrinsic normalized error will
remain within $[-1,1]$ for most bins, and we can see that error
introduced by the NN similarly does not exceed this magnitude.

We note that networks used to obtain these results were trained in a
few CPU hours and thereafter can make an accurate determination of
whether a GRB will be detected by Swift in just a few microseconds,
instead of the minutes of computation time required by the full
simulated Swift trigger pipeline.

\subsection{Dimensionality reduction: compressing galaxy images}\label{sec:MDMautoencoder}

In astronomical data analysis, the raw data set often contains a great
deal of redundant information. Simply put, there are usually many more
pixels in an astronomical image than there are distinct features to be
identified in the object being imaged, so that not all pixels are
independent measures of structure. If we are able to compress the data
by removing these redundancies and instead quantify only the distinct
features present, then one can be more efficient in subsequent
analyses. One way of finding these features in the data and performing
compression -- and denoising -- is through the use of autoencoders. As
we have already shown, autoencoders are able to represent non-linear
features in a data set and reduce the number of variables used to
describe it to a value closer to the intrinsic dimensionality of the
data.

\subsubsection{Image compression and denoising}\label{sec:MDMcompress}

Our previous analysis, presented in Section~\ref{sec:MDM}, for
measuring galaxy ellipticities from the MDM Challenge data set images
provides a good example. The original galaxy images each contain
$2304$ pixels, but these are clearly not all independent
measurements. Additionally, even the cropped star images each contain
$576$ non-independent pixels. With an autoencoder, we can compress
both of these images by two orders of magnitude, from thousands of
input variables to just tens, from which one can then measure
ellipticities.

To perform the compression, autoencoders with a single hidden layer
were trained, since these images contain relatively few features and
simpler networks require less time to train. Pre-training was used in
all examples. The input data were the $2880$ pixels of each galaxy and
its accompanying (cropped) star. Since the noise in each pixel is
Poissonian, we report in Table~\ref{tab:MDMautopred} not only the RMSE
on the autoencoder outputs, but also the RMSE normalised by the
original pixel value. The values listed in Table~\ref{tab:MDMautopred}
correspond to the mean values obtained for the collection of images
comprising the MDM training data set.
\begin{table}
\caption{The mean RMSE values for autoencoder reconstructions of
  galaxy and cropped star images from the MDM Challenge data
  set. Pixel values range from $0$ to $255$. The normalized RMSE has
  had the errors divided by the square root of the original pixel
  value that was to be reconstructed.}
\label{tab:MDMautopred}
\begin{center}
\begin{tabular}{ccc}
\hline
Hidden nodes & RMSE & Norm. RMSE \\ \hline 
$1$ & $9.09$ & $0.955$ \\
$2$ & $8.83$ & $0.936$ \\
$3$ & $8.66$ & $0.922$ \\
$5$ & $8.50$ & $0.909$ \\
$10$ & $8.44$ & $0.904$ \\
$30$ & $8.43$ & $0.903$ \\
$50$ & $8.42$ & $0.901$ \\
$100$ & $8.38$ & $0.897$ \\
\hline
\end{tabular}
\end{center}
\end{table}
The values obtained from one image to another are mutually consistent,
but when the autoencoder has $10$ or more nodes in the hidden layer
the image-to-image variations are significantly larger than the
differences between the means. One sees from
Table~\ref{tab:MDMautopred} that only a slight improvement is observed
when using 10 or more hidden nodes. Indeed, taking into account
image-to-image variation, the predictions by the larger networks are
statistically indistinguishable from those of the smaller ones. We can
therefore represent the images well with just $10$ feature values. If
we consider the original construction of the images, each galaxy can
be represented by $4$ parameters (two ellipticities, a position angle
and an amplitude) and each star by $2$ parameters (a radius for the
point-spread function and an amplitude). This means that, without
noise, only $6$ parameters are needed to describe the images
completely. This is reflected by the ability of our autoencoder to
perform the `majority' of the fit with just $5$ features. Additional
features produce more marginal decreases in the RMSE as they are now
fine-tuning and/or fitting for the noise in the data.

\begin{figure}
\begin{center}
\includegraphics[width=0.99\columnwidth]{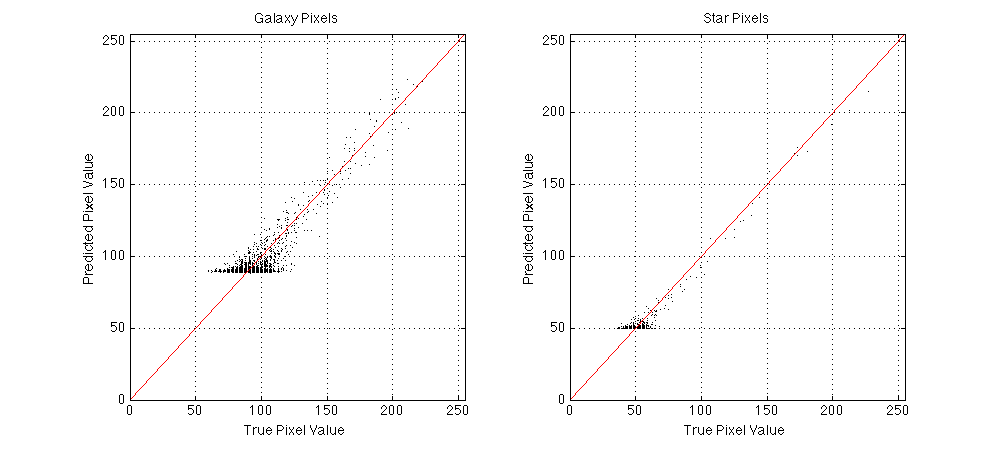}
\caption{A pixel-by-pixel comparison for a typical galaxy/star image
  pair between original and autoencoder reconstructed images. We can
  see the larger errors occurring at smaller true pixel values
  associated with the background. This used an autoencoder with a
  single hidden layer of $10$ nodes.}
\label{fig:MDMpixelcomparison}
\end{center}
\end{figure}

\begin{figure}
\begin{center}
\includegraphics[width=0.99\columnwidth]{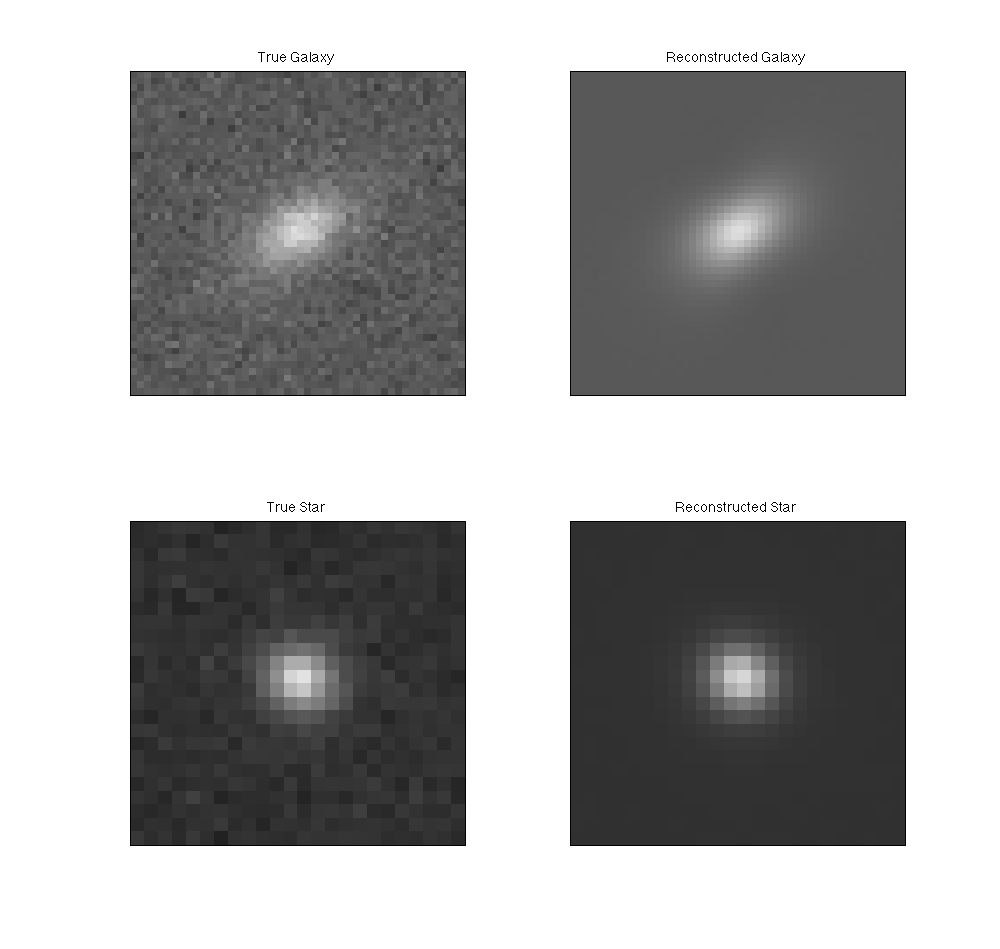}
\caption{Comparison of original (input) and reconstructed (output)
  galaxy and star images for an autoencoder with a single hidden layer
  of $10$ nodes. This example is the same as that shown in
  Figure~\ref{fig:MDMpixelcomparison}}
\label{fig:MDMautoreconstruct}
\end{center}
\end{figure}

By looking at pixel-to-pixel comparisons, one finds that a large part
of the error is coming from the fainter pixels, which are, for the
most part, distinctly external to the galaxy. Therefore, the galaxy
itself is being described even more accurately than the numbers
presented indicate. A plot of this comparison for a typical
galaxy/star pair is shown in Figure~\ref{fig:MDMpixelcomparison}, with
the corresponding original (input) and reconstructed (output) images
shown in Figure~\ref{fig:MDMautoreconstruct}.

\begin{figure}
\begin{center}
\subfigure[]{\includegraphics[width=0.99\columnwidth]{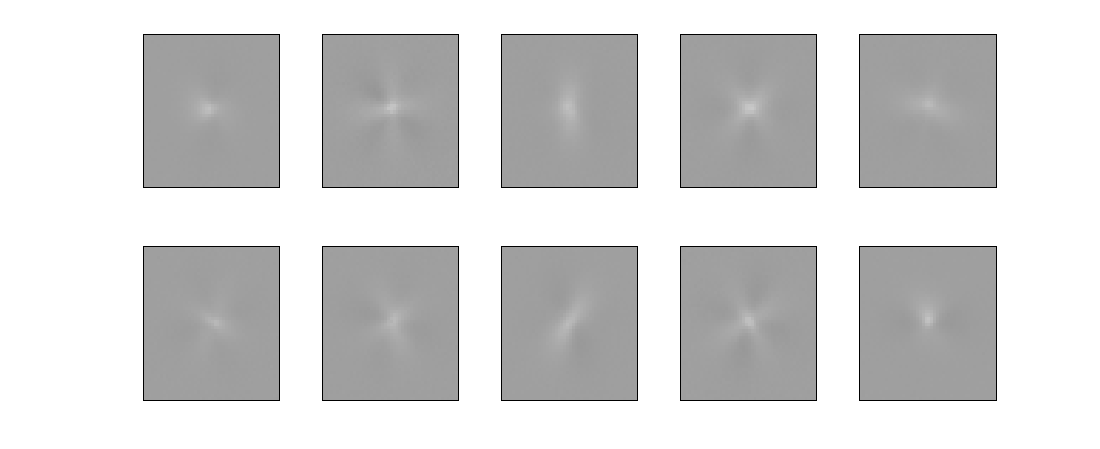}}
\subfigure[]{\includegraphics[width=0.99\columnwidth]{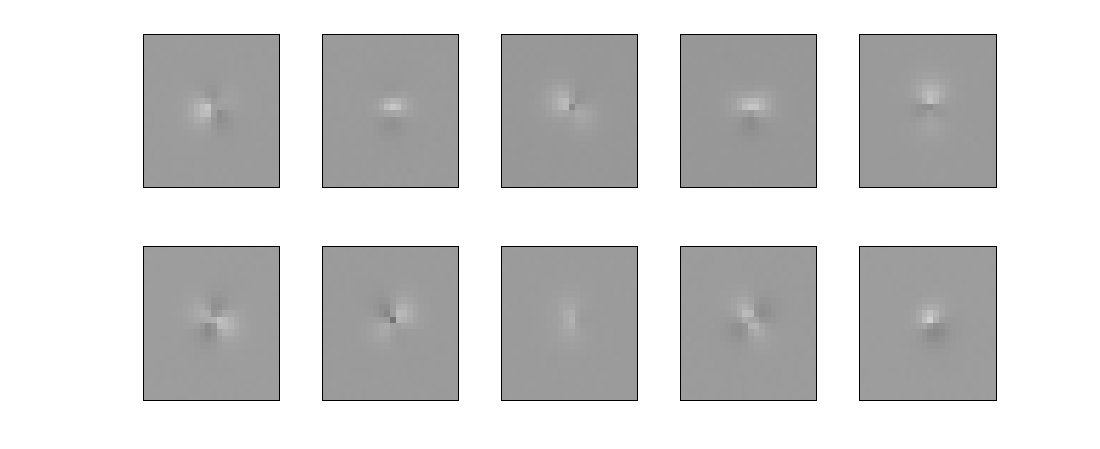}}
\caption{Features vectors obtained by decoding $(1,0,0,\ldots)$,
  $(0,1,0,\ldots)$, etc. in the central layer of an autoencoder with
  $10$ central layer nodes. Shown are the extracted (a) galaxy and (b)
  star features. The greyscale has been reversed on the actual values.}
\label{fig:MDMfeaturevectors}
\end{center}
\end{figure}

We can illustrate the nature of the feature vectors constructed by the
network by decoding the central layer values $(1,0,0,\ldots)$,
$(0,1,0,\ldots)$, etc. of our autoencoder, which has 10 hidden nodes, to obtain
the corresponding 10 output images. We plot these images in
Figure~\ref{fig:MDMfeaturevectors} for the galaxy/star example shown
in Figure~\ref{fig:MDMautoreconstruct}. Some shape features can be
clearly seen, although the greyscale has been reversed for these images to see
them as brighter structures. This reversal can be accounted for by the network
assigning the original features negative weights and a positive bias.

\subsubsection{Estimating ellipticities with compressed data}\label{sec:MDMcompressedmeasure}

Having trained the autoencoders, we now investigate using the
compressed feature values, rather than original images, to determine
the ellipticities of the galaxies. Since the number of inputs has been
decreased from $2880$ to $10$--$100$, we can use more and larger
hidden layers in our regression network and the analysis will still
take less time to run. Results on training regression networks with
many different configurations are given in
Table~\ref{tab:MDMcompressedmeasureresults}.
\begin{table}
\caption{RMSE values for galaxy ellipticity predictions using the
  compressed feature values as inputs. Various compressions and
  regression network sizes were used.}
\label{tab:MDMcompressedmeasureresults}
\begin{center}
\begin{tabular}{ccc}
\hline
\# Features & Hidden layers & RMSE \\ \hline 
\multirow{5}{*}{5} & $10$ & $0.022316$ \\
 & $30$ & $0.022534$ \\
 & $100$ & $0.022015$ \\
 & $30+10$ & $0.025472$ \\
 & $100+30$ & $0.025165$ \\ \hline
\multirow{5}{*}{10} & $10$ & $0.016802$ \\
 & $30$ & $0.016237$ \\
 & $100$ & $0.016296$ \\
 & $30+10$ & $0.017028$ \\
 & $100+30$ & $0.017869$ \\ \hline
\multirow{5}{*}{30} & $10$ & $0.016559$ \\
 & $30$ & $0.016927$ \\
 & $100$ & $0.016608$ \\
 & $30+10$ & $0.017312$ \\
 & $100+30$ & $0.017351$ \\ \hline
\multirow{5}{*}{50} & $10$ & $0.017056$ \\
 & $30$ & $0.017056$ \\
 & $100$ & $0.016769$ \\
 & $30+10$ & $0.016492$ \\
 & $100+30$ & $0.016629$ \\ \hline
\multirow{5}{*}{100} & $10$ & $0.019459$ \\
 & $30$ & $0.019561$ \\
 & $100$ & $0.019596$ \\
 & $30+10$ & $0.019750$ \\
 & $100+30$ & $0.019949$ \\ \hline
\hline
\end{tabular}
\end{center}
\end{table}

These results show that the extra information given to the regression
networks trained on $100$ feature values from the autoencoder acted as
a disadvantage in predicting the galaxy ellipticities. For networks
trained on $50$ or $30$ features, however, the accuracies of the
predicted ellipticities were better even than those obtained using the
full original images in some cases. This demonstrates the power of
being able to eliminate redundant information and noise, and thereby
improve the accuracy of the analysis. We also observe that adding
unnecessary complexity to the NN structure makes it more difficult for
the algorithm to find the global maximum.

This same method for dimensionality reduction -- which also eliminates
noise -- before performing measurements can clearly be applied to a
wide range of other astronomical applications. Examples include
classification of supernovae by type, or measurements of galaxies and
stars by their spectra.

\section{Conclusions}\label{sec:NNdiscuss}

We have described an efficient and robust neural network training
algorithm, called {\sc SkyNet}, which we have now made freely
available for academic purposes. This generic tool is capable of
training large and deep feed-forward networks, including autoencoders,
and may be applied to supervised and unsupervised machine learning
tasks in astronomy, such as regression, classification, density
estimation, clustering and dimensionality reduction. {\sc SkyNet}
employs a combination of (optional) pre-training followed by iterative
refinement of the network parameters using a
regularised variant of Newton's optimisation algorithm
that incorporates second-order derivative information without the need
even to compute or store the Hessian matrix. Linear and sigmoidal
activation functions are provided for the user to choose between.
{\sc SkyNet} adopts convergence criteria that naturally prevent overfitting,
and it also includes a fast algorithm for estimating the accuracy of network
outputs.

We first demonstrate the capabilities of {\sc SkyNet} on toy examples
of regression, classification, and dimensionality reduction using
autoencoder networks, and then apply it to the classic
machine learning problem of handwriting classification for determining
digits from the MNIST database. In an astronomical context, {\sc
  SkyNet} is applied to: the regression problem of measuring the
ellipticity of noisy and convolved galaxy images in the Mapping Dark
Matter Challenge; the classification problem of identifying gamma-ray
bursters that are detectable by the Swift satellite; and the
dimensionality reduction problem of compressing and denoising images
of galaxies. In each case, the straightforward use of {\sc SkyNet}
produces networks that perform the desired task quickly and
accurately, and typically achieve results that are competitive with
machine learning approaches that have been tailored to the required
task.

Future development of {\sc SkyNet} will expand upon many of the current features
and introduce new ones. We are working to include more activation functions
(e.g. $\tanh$, softsign, and rectified linear), pooling of nodes, convolutional NNs,
diversity in outputs (i.e. mixing regression and classification), and more robust
support of recursive NNs. This is all in addition to further improving the speed and
efficiency of the training algorithm itself. However, {\sc SkyNet} in its current
state is already a useful tool for performing machine learning in astronomy.

\section*{Acknowledgments}\label{sec:Acknowledgments}
The authors thank John Skilling for providing very useful advice in
the early stages of algorithm development. We also thank Amy Lien for
providing the data used in Seciton~\ref{sec:GRB}. This work utilized
three different high-performance computing facilities at different
times: initial work was performed on COSMOS VIII, an SGI Altix UV1000
supercomputer, funded by SGI/Intel, HEFCE and PPARC, and the authors
thank Andrey Kaliazin for assistance; early work also utilized the
Darwin Supercomputer of the University of Cambridge High Performance
Computing Service (\texttt{http://www.hpc.cam.ac.uk/}), provided by
Dell Inc. using Strategic Research Infrastructure Funding from the
Higher Education Funding Council for England; later work utilised the
Discover system of the NASA Center for Climate Simulation at NASA
Goddard Space Flight Center. PG is currently supported by a NASA
Postdoctoral Fellowship from the Oak Ridge Associated Universities and
completed a portion of this work while funded by a Gates Cambridge
Scholarship at the University of Cambridge. FF is supported by a
Research Fellowship from the Leverhulme and Newton Trusts.

\setlength{\labelwidth}{0pt}

\label{lastpage}


\begin{thebibliography}{1}
\bibitem[\protect\citeauthoryear{Andreon et al.}{1999}]{Andreon1999} Andreon S., Gargiulo G., Longo G., Tagliaferri R., \& Capuano N., 1999, arXiv:astro-ph/9906099
\bibitem[\protect\citeauthoryear{Andreon et al.}{2000}]{Andreon2000} Andreon S., Gargiulo G., Longo G., Tagliaferri R., \& Capuano N., 2000, MNRAS, 319, 700--716
\bibitem[\protect\citeauthoryear{Auld et al.}{2008}]{cosmonet1} Auld T., Bridges M., Hobson M.P., Gull S.F., 2008, MNRAS, 376, L11
\bibitem[\protect\citeauthoryear{Auld, Bridges \& Hobson}{2008}]{cosmonet2} Auld T., Bridges M., Hobson M.P., 2008, MNRAS, 387, 1575
\bibitem[\protect\citeauthoryear{Ball \& Brunner}{2010}]{Ball2010} Ball N.M., Brunner R.J., 2010, Int.~J.~Mod.~Phys., 19, 1049
\bibitem[\protect\citeauthoryear{Begstra et al.}{2009}]{Bergstra2009} Bergstra J., Desjardins G., Lamblin P., \& Bengio Y., 2009, Technical Report 1337, D\'{e}partement d’Informatique et de Recherche Op\'{e}rationnelle, Universit\'{e} de Montr\'{e}al.
\bibitem[\protect\citeauthoryear{Bertin and Arnouts}{1996}]{SExtractor} Bertin E., Arnouts S., 1996, A\&AS Supplement, 117, 393
\bibitem[\protect\citeauthoryear{Bridges et al.}{2011}]{coverage} Bridges M., Cranmer K., Feroz F., Hobson M.P., Ruiz de Austri R., Trotta R., 2011, JHEP, 03, 012
\bibitem[\protect\citeauthoryear{Bonnett}{2013}]{Bonnett2013} Bonnett C., 2013, arXiv:1312.1287 [astro-ph.CO]
\bibitem[\protect\citeauthoryear{Carreira-Perpignan \& Hinton}{2005}]{contrastdiv} Carreira-Perpignan M. A. \& Hinton. G. E., 2005, Proceedings of the Tenth International Workshop on Artificial Intelligence and Statistics, eds. Cowell R. G. \& Ghahramani Z., 33--40
\bibitem[\protect\citeauthoryear{Ciresan et al.}{2010}]{MNISTdeepsoln} Ciresan D. C., Meier U., Gambardella L. M., \& Schmidhuber J., 2010, Neural Comput., 22, 3207--3220
\bibitem[\protect\citeauthoryear{Cybenko}{1989}]{UnivApprox2} Cybenko G., 1989, Mathematics of Control, Signals, and Systems, 2, 303--314
\bibitem[\protect\citeauthoryear{Erhan et al.}{2010}]{Erhan2010} Erhan D. et al., 2010, Journal of Machine Learning Research, 11, 625--660
\bibitem[\protect\citeauthoryear{Fawcett}{2006}]{Fawcett2006} Fawcett T., 2006, Pattern Recogn. Lett., 27, 861
\bibitem[\protect\citeauthoryear{Fendt \& Wandelt}{2007}]{pico} Fendt W.A., Wandelt B.D., 2007, ApJ, 654, 2
\bibitem[\protect\citeauthoryear{Feroz \& Hobson}{2008}]{multimodalNS} Feroz F., Hobson M.P., 2008, MNRAS, 384, 449
\bibitem[\protect\citeauthoryear{Feroz, Hobson \& Bridges}{2009}]{multinest} Feroz F., Hobson M.P., Bridges M., 2009, MNRAS, 398, 1601
\bibitem[\protect\citeauthoryear{Feroz et al.}{2013}]{multinest3} Feroz F., Hobson M. P., Cameron E., \& Pettitt A. N., 2013, arXiv:1306.2144 [astro-ph.IM]
\bibitem[\protect\citeauthoryear{Feroz, Marshall \& Hobson}{2008}]{Feroz2008} Feroz F., Marshall P. J., Hobson M. P., 2008, arXiv:0810.0781 [astro-ph]
\bibitem[\protect\citeauthoryear{Fynbo et al.}{2009}]{Fynbo2009} Fynbo J. et al., 2009, ApJS, 185, 526
\bibitem[\protect\citeauthoryear{Gehrels et al.}{2004}]{Gehrels2004} Gehrels N. et al., 2004, ApJ, 611, 1005
\bibitem[\protect\citeauthoryear{Geva \& Sitte}{1992}]{GevaSitte} Geva S., Sitte J., IEEE, 3, 621
\bibitem[\protect\citeauthoryear{Glorot \& Bengio}{2010}]{GlorotBengio2010} Glorot X. \& Bengio Y, 2010, Proceedings of the Thirteenth International Conference on Artificial Intelligence and Statistics, Journal of Machine Learning Research, eds. Teh Y. W. \& Titterington M., 249--256
\bibitem[\protect\citeauthoryear{Glorot et al.}{2011}]{Glorot2011ReLU} Glorot X., Bordes A., \& Bengio Y., 2011, Proceedings of the Fourteenth International Conference on Artificial Intelligence and Statistics, Journal of Machine Learning Research, eds. Gordon G. \& Dunson D., 315--323
\bibitem[\protect\citeauthoryear{Graff et al.}{2012}]{bambi}
Graff P., Feroz F., Hobson M.P., Lasenby A.N., 2012, MNRAS, 421, 169
\bibitem[\protect\citeauthoryear{Gull \& Skilling}{1999}]{MemSys} Gull S.F. \& Skilling J., 1999, Quantified Maximum Entropy:  MemSys 5 Users' Manual. Maximum Entropy Data Consultants Ltd. Bury St. Edmunds, Suffolk, UK. \texttt{http://www.maxent.co.uk/}
\bibitem[\protect\citeauthoryear{Hinton, Osindero, \& Teh}{2006}]{HintonFastRBMmethod} Hinton G.E., Osindero S., \& Teh Y.-W., 2006, Neural Comput., 18, 1527--1554
\bibitem[\protect\citeauthoryear{Hinton \& Salakhutdinov}{2006}]{Hinton_AE_Science} Hinton G.E. \& Salakhutdinov R.R., 2006, Science, 313, 504-507
\bibitem[\protect\citeauthoryear{Hobson et al.}{1998}]{Hobson98} Hobson M. P., Jones A. W., Lasenby A. N., \& Bouchet F. R., 1998, MNRAS, 300, 1--29
\bibitem[\protect\citeauthoryear{Hornik, Stinchcombe \& White}{1990}]{UnivApprox} Hornik K., Stinchcombe M. \& White H., 1990, Neural Networks, 3, 359
\bibitem[\protect\citeauthoryear{Hyv\"arinen \& Oja}{2000}]{Hyvarinen2000} Hyv\"arinen A., Oja E., 2000, Neural Networks, 13, 411
\bibitem[\protect\citeauthoryear{Karpenka, Feroz \& Hobson}{2013}]{karpenka} Karpenka N.V., Feroz F., Hobson M.P., 2013, MNRAS, 429, 1278--1285
\bibitem[\protect\citeauthoryear{Kendall}{1957}]{kendall} Kendall M.G., 1957, A course in multivariate analysis. Griffin, London
\bibitem[\protect\citeauthoryear{Kitching et al.}{2011}]{GREAT10} Kitching T. et al., 2011, Annals of Applied Statistics, 5, 2231--2263
\bibitem[\protect\citeauthoryear{Kitching et al.}{2012}]{MDMresults} Kitching T. et al., 2012, New Astronomy Reviews, submitted
\bibitem[\protect\citeauthoryear{LeCun et al.}{1998}]{MNISTlecun98} LeCun Y., Bottou L., Bengio Y., \& Haffner P., 1998, Proc. of the IEEE, 86, 2278--2324
\bibitem[\protect\citeauthoryear{Lien et al.}{2012}]{LienGRB} Lien A., Sakamoto T., Gehrels N., Palmer D., Graziani C., 2012, Proceedings of the International Astronomical Union, 279, 347
\bibitem[\protect\citeauthoryear{Longo, Tagliaferri \& Andreon}{2001}]{LTA2001} Longo G., Tagliaferri R., \& Andreon S., 2001, Mining the Sky: Proceedings of the MPA/ESO/MPE Workshop, eds. Banday A. J., Zaroubi S., Bartelmann M., 379--385
\bibitem[\protect\citeauthoryear{MacKay}{1992}]{MacKay_BayesianInterp} MacKay D.J.C., 1992, Neural Computation, 4, 415--447
\bibitem[\protect\citeauthoryear{MacKay}{1995}]{NNprederror} MacKay D.J.C., 1995, Network: Computation in Neural Systems, 6, 469
\bibitem[\protect\citeauthoryear{MacKay}{2003}]{MacKay_ITILA} MacKay D.J.C, 2003, Information Theory, Inference, and Learning Algorithms. Cambridge Univ. Press. \texttt{www.inference.phy.cam.ac.uk/mackay/itila/}
\bibitem[\protect\citeauthoryear{Mandic \& Chambers}{2001}]{Mandic2001} Mandic D., Chambers J., 2001, Recurrent Neural Networks for Prediction: Learning Algorithms, Architectures and Stability. Wiley, New York.
\bibitem[\protect\citeauthoryear{Martens}{2010}]{HessianFree} Martens J., 2010, in F\"{u}rnkranz J., Joachims T., eds, Proc. 27th Int. Conf. Machine Learning. Omnipress, Haifa, p. 735
\bibitem[\protect\citeauthoryear{Murtagh}{1991}]{murtagh} Murtagh F., 1991, Neural Comput., 2, 183
\bibitem[\protect\citeauthoryear{Pascanu \& Bengio}{2013}]{Pascanu2013} Pascanu R. \& Bengio Y., 2013, arXiv:1301.3584 [cs.LG]
\bibitem[\protect\citeauthoryear{Pearlmutter}{1994}]{Pearlmutter} Pearlmutter B.A., 1994, Neural Comput., 6, 147
\bibitem[\protect\citeauthoryear{Sanger}{1989}]{sanger} Sanger T.D., 1989, Neural Networks, 2, 459
\bibitem[\protect\citeauthoryear{Schraudolph}{2002}]{Schraudolph} Schraudolph N.N., 2002, Neural Comput., 14, 1723
\bibitem[\protect\citeauthoryear{Serra-Ricart et al.}{1993}]{AstroNN} Serra-Ricart M., Calbet X., Garrido L., \& Gaitan V., 1993, AJ, 106, 1685
\bibitem[\protect\citeauthoryear{Skilling}{2004}]{Skilling} Skilling J., 2004, AIP Conference Series, 735, 395
\bibitem[\protect\citeauthoryear{Tagliaferri et al.}{2003a}]{Tagliaferri2003a} Tagliaferri R. et al., 2003a, Neural Networks, 16, 297
\bibitem[\protect\citeauthoryear{Tagliaferri et al.}{2003b}]{Tagliaferri2003b} Tagliaferri R., Longo G., Andreon S., Capozziello S., Donalek C., \& Giordano G., 2003b, Neural Nets: 14th Italian Workshop on Neural Nets, eds. Apolloni B., Marinaro M., \& Tagliaferri R., 226--234
\bibitem[\protect\citeauthoryear{Wanderman \& Piran}{2010}]{Wanderman2010} Wanderman D., Piran T., 2010, MNRAS, 406, 1944
\bibitem[\protect\citeauthoryear{Way et al.}{2012}]{Way2012} Way M.J., Scargle J.D., Ali K.M., Srivastava A.N., 2012, Advances in Machine Learning and Data Mining for Astronomy. CRC Press. 
\end{thebibliography}
\end{document}